\documentclass[10pt]{article}

\usepackage{hyperref}
\usepackage[utf8]{inputenc} 
\usepackage[T1]{fontenc}    
\usepackage{hyperref}       
\usepackage{url}            
\usepackage{booktabs}       
\usepackage{amsfonts}       
\usepackage{nicefrac}       
\usepackage{microtype}      
\newcommand{\red}[1]{{\color{red} #1}}
\usepackage{overpic} 
\usepackage{amsmath,amssymb,bbm,stmaryrd,mathrsfs,url}
\newcommand{\nat}{\natural}
\usepackage{mwe,subfig,color,xcolor}

\newcommand{\R}{\mathbb{R}}
\newcommand{\C}{\mathbb{C}}

\newcommand{\E}{\mathbb{E}}

\newcommand{\vct}[1]{\boldsymbol{#1}}
\newcommand{\mtx}[1]{\boldsymbol{#1}}

\newcommand{\<}{\langle}
\renewcommand{\>}{\rangle}

\newcommand{\trace}{\operatorname{Tr}}

\newcommand{\set}[1]{\mathcal{#1}}

\DeclareMathOperator*{\minimize}{\text{minimize}}

\newcommand{\va}{\vct{a}}
\newcommand{\vb}{\vct{b}}
\newcommand{\vc}{\vct{c}}

\newcommand{\vg}{\vct{g}}
\newcommand{\vh}{\vct{h}}

\newcommand{\vm}{\vct{m}}

\newcommand{\vu}{\vct{u}}
\newcommand{\vv}{\vct{v}}
\newcommand{\vw}{\vct{w}}
\newcommand{\vx}{\vct{x}}
\newcommand{\vy}{\vct{y}}
\newcommand{\vz}{\vct{z}}

\newcommand{\vxi}{\vct{\xi}}
\newcommand{\vzero}{\vct{0}}

\newcommand{\mB}{\mtx{B}}
\newcommand{\mC}{\mtx{C}}

\newcommand{\mF}{\mtx{F}}
\newcommand{\mG}{\mtx{G}}
\newcommand{\mH}{\mtx{H}}
\newcommand{\mI}{\mtx{I}}

\newcommand{\mM}{\mtx{M}}

\newcommand{\mV}{\mtx{V}}
\newcommand{\mW}{\mtx{W}}
\newcommand{\mX}{\mtx{X}}

\newcommand{\setB}{\set{B}}

\newcommand{\setH}{\set{H}}
\newcommand{\setI}{\set{I}}
\newcommand{\setJ}{\set{J}}

\newcommand{\setL}{\set{L}}

\newcommand{\setN}{\set{N}}

\newcommand{\setP}{\set{P}}
\newcommand{\setQ}{\set{Q}}
\newcommand{\setR}{\set{R}}

\newcommand{\yl}{y_\ell}

\newcommand{\bl}{\vb_\ell}

\newcommand{\cl}{\vc_\ell}

\newcommand{\dM}{\delta \mM}
\newcommand{\dH}{\delta \mH}

\newcommand{\blt}{\bl^*}
\newcommand{\clt}{\cl^*}

\newcommand{\PP}{\mathbb{P}}

\newcommand{\Th}{T_{\tilde{h}}}
\newcommand{\Tm}{T_{\tilde{m}}}
\newcommand{\Thp}{T_{\tilde{h}}^\perp}
\newcommand{\Tmp}{T_{\tilde{m}}^\perp}
\newcommand{\PTh}{\setP_{\Th}}
\newcommand{\PTm}{\setP_{\Tm}}

\newcommand{\ba}[1]{{\va_{{#1},\ell}}}
\newcommand{\ind}{\mathbb{I}}
\newcommand{\balpha}{\boldsymbol{\alpha}}
\DeclareMathOperator*{\argmin}{arg\,min}
\allowdisplaybreaks

\makeindex

\newtheorem{lemma}{Lemma}
\newtheorem{corollary}{Corollary}

\newtheorem{theorem}{Theorem}

\begin{document}

\title{\vspace{-2cm}\bf{Simultaneous Phase Retrieval and Blind Deconvolution via Convex Programming}}
\author{Ali Ahmed\thanks{Department of Electrical Engineering, Information Technology University, Lahore, Pakistan; Email: {\tt ali.ahmed@itu.edu.pk}}\\ \and  Alireza Aghasi\thanks{Department of Business Analytics, Georgia State University, Atlanta, GA; Email: {\tt aaghasi@gsu.edu}}\\ \and Paul Hand\thanks{Department of Mathematics and Khoury College of Computer Sciences, Northeastern University, Boston, MA; Email: {\tt p.hand@northeastern.edu}}\\}

\date{}    


\maketitle

\begin{abstract}
We consider the task of recovering two real or complex $m$-vectors from phaseless Fourier measurements of their circular convolution.  Our method is a novel convex relaxation that is based on a lifted matrix recovery formulation that allows a nontrivial convex relaxation of the bilinear measurements from convolution.    We prove that if  the two signals belong to known random subspaces of dimensions $k$ and $n$, then they can be recovered up to the inherent scaling ambiguity with $m  \gg (k+n) \log^2 m$  phaseless measurements.  Our method provides the first theoretical recovery guarantee for this problem by a computationally efficient algorithm and does not require a solution estimate to be computed for initialization. Our proof is based on Rademacher complexity estimates.  Additionally, we provide an alternating direction method of multipliers (ADMM) implementation and provide numerical experiments that verify the theory.
\end{abstract}

\textbf{Keywords:} Hyperbolic Constraints, Blind Deconvolution, Phase Retrieval, Convex Analysis, Rademacher Complexity

\section{Introduction}

This paper considers recovery of two unknown signals (real- or complex-valued) from the magnitude only measurements of their convolution.  Let $\vw$, and $\vx$ be vectors residing in $\setH^m$, where $\setH$ denotes either $\R$, or $\C$. Moreover, denote by $\mF$ the DFT matrix with entries $F[\omega,t] =\tfrac{1}{\sqrt{m}} \mathrm{e}^{-j2\pi \omega t/m}, ~ 1 \leq \omega, t \leq m.$ We observe the phaseless Fourier coefficients of the circular convolution $\vw \circledast \vx$ of $\vw$, and $\vx$ 
\begin{align}\label{eq:measurements}
\tilde{\vy} = |\mF(\vw \circledast \vx)|,
\end{align}
where $|\vz|$ returns the element wise absolute value of the vector $\vz$. We use $\tilde{\vy}$ to denote  noiseless measurements, and reserve the notation $\vy$ for more general noisy measurements. We are interested in recovering $\vw$ and $\vx$ from the phaseless measurements $\tilde{\vy}$ or $\vy$ of their circular convolution. In other words, the problem concerns blind deconvolution of two signals from phaseless measurements. The problem can also be viewed as identifying the structural properties on $\vw$ such that its convolution with the signal/image of interest $\vx$ makes the phase retrieval of a signal $\vx$ well-posed. 
Since $\vw$ and $\vx$ are both unknown, and in addition, the measurements are phaseless, the inverse problem becomes severly ill-posed as many pairs of $\vw$ and $\vx$ correspond to the same $\vy$. We show that this non-linear problem can be efficiently solved, under Gaussian measurements, using a semidefinite program and also theoretically prove this assertion. We also propose a heuristic approach to solve the proposed semidefinite program computationally efficiently. Numerical experiments show that, using this algorithm, one can successfully recover a blurred image from the magnitude only measurements of its Fourier spectrum. 

Phase retrieval has been of continued interest in the fields of signal processing, imaging, physics, computational science, etc. Perhaps, the single most important context in which phase retrieval arises is the X-ray crystallography \cite{harrison1993phase,millane1990phase}, where the far-field pattern of X-rays scattered from a crystal form a Fourier transform of its image, and it is only possible to measure the intensities of the electromagnetic radiation. However, with the advancement of imaging technologies, the phase retrieval problem continues to arise in several other imaging modalities such as diffraction imaging \cite{bunk2007diffractive}, microscopy \cite{miao2008extending}, and astronomical imaging \cite{fienup1987phase}. In the imaging context, the result in this paper would mean that if rays are convolved with a \textit{generic} pattern (either man made or naturally arising due to propagation of light through some unknown media) $\vw$ prior to being scattered/reflected from the object, the image of the object can be recovered from the Fourier intensity measurements later on. As is well known from Fourier optics \cite{goodman2008introduction}, the convolution of a visible light with a generic pattern can be implemented using a lens-grating-lens setup. 

Despite recent advances in theoretical understanding of  phase retrieval \cite{candes2013phaselift,candes2015phasecoded}, the application to actual problems such as crystallography remains challenging owing partly to the simplistic mathematical models that may not fully capture the actual physical problem at hand. Our comparatively more complex model in \eqref{eq:measurements} more elaborately encompasses structure in actual physical problems, for example, crystallography, where due to the natural periodic arrangement of a crystal structural unit, the observed electron density function of the crystal exactly takes the form \eqref{eq:measurements}; for details, see, Section 2 of \cite{elser2017benchmark}.

Blind deconvolution is a fundamental problem in signal processing, communications, and in general system theory. Visible light communication has been proposed as a standard in 5G communications for local area networks \cite{azhar2013gigabit,retamal20154,azhar2010demonstration}. Propagation of information carrying light through an unknown communication medium is modeled as a convolution. The channel is unknown and at the receiver it is generally difficult to measure the phase information in the propagated light. The result in this paper says that the transmitted signal can be blindly deconvolved from the unknown channel using the Fourier intensity measurements of the light only. The reader is referred to the first section of the supplementary note for a detailed description of the visible light communication and its connection to our formulation. 

\textbf{Main Contributions.} In this paper, we study the combination of two important and notoriously challenging signal recovery problems: phase retrieval and blind deconvolution.  We introduce a novel convex formulation that is possible because the algebraic structure from lifting resolves the bilinear ambiguity just enough to permit a nontrivial convex relaxation of the measurements.  The strengths of our approach are that it allows a novel convex program that is the first to provably permit recovery guarantees with optimal sample complexity for the joint task of phase retrieval and blind deconvolution when the signals belong to known random subspaces.  Additionally, unlike many recent convex relaxations and nonconvex approaches, our approach does not require an initialization or estimate of the true solution in order to be stated or solved.  While our convex formulation is presented in a lifted domain (with increased dimensionality), in implementing the convex problem, we have been able to use some recent results in Burer-Monteiro-type approaches and perform the optimization in a factored space (solving a series of nonconvex programs which are guaranteed to land on the global minima).


Finally, an earlier version of this paper with only the exact recovery result form noiseless measurements \label{eq:noiseless-measurments} appeared in \cite{AAH2018Blind} by the same authors. This paper extends the previous result to more general noisy measurements with a significantly modified proof. Moreover, the implementation in \cite{AAH2018Blind} was performed in a lifted domain and the proposed scheme required iterative projections onto the positive semidefinite cone, which was computationally prohibitive for large scale problems. By considering a different way of modeling the optimization problem, in Section \ref{sec:Implementation} we present a more efficient algorithm, which is solved in a factored space using a Burer-Monteiro-type approach. This makes our implementation applicable to a much larger class of problems.  

\subsection{Observations in Matrix Form}

The phase retrieval, and blind deconvolution problem has been extensively studied in signal processing community in recent years \cite{candes2015phase,ahmed2014blind} by lifting the unknown vectors to a higher dimensional matrix space formed by their outer products. The resulting rank-1 matrix is recovered using nuclear norm as a convex relaxation of the non-convex rank constraint. Recently, other forms of convex relaxations have been proposed \cite{bahmaniphaseretrieval,goldstein2018phasemax,aghasi2017branchhull,aghasi2017convex,aghasi2018convex} that solve both the problems in the native (unlifted) space leading to computationally efficiently solvable convex programs. This paper handles the non-linear convolutional phase retrieval problem by lifting it into a bilinear problem. The resulting problem, though still non-convex, gives way to an effective convex relaxation that provably recovers $\vw$ and $\vx$ exactly. 

We consider the problem of recovering $(\vw^\nat, \vx^\nat) \in \setH^{\ell} \times \setH^{\ell}$ from measurements of the form \eqref{eq:measurements}.
It is clear that uniquely recovering $\vw^\nat$ and $\vx^\nat$, even up to the global bilinear abiguity, is not possible without extra knowledge or information about the problem. We will address the problem under the broad and generally applicable structural assumptions that both $\vw^\nat$ and $\vx^\nat$ are members of known subspaces of $\setH^m$. This means that $\vw^\nat$ and $\vx^\nat$ can be parameterized in terms of unknown lower dimensional vectors $\vh^\nat \in \setH^k$ and $\vm^\nat \in \setH^n$, respectively, as follows
\begin{align}\label{eq:subspace-constraints}
\vw^\nat = \mB\vh^\nat, \ \vx^\nat = \mC \vm^\nat, 
\end{align}
where $\mB \in \setH^{m \times k}$, and $\mC \in \setH^{m \times n}$ are known matrices whose columns span the subspaces in which $\vw^\nat$ and $\vx^\nat$ belong, respectively.  Since the circular convolution operator diagonalizes in the Fourier domain, noiseless measurements become
\begin{align}\label{eq:noiseless-measurements}
\tilde{\vy} = \tfrac{1}{\sqrt{m}}|\hat{\mB}\vh^\nat \odot \hat{\mC}\vm^\nat|,
\end{align}
where $\hat{\mB} = \sqrt{m}\mF\mB$, $\hat{\mC} = \sqrt{m}\mF\mC$, and $\odot$ represents the the Hadamard product.  Denoting by $\bl$ and $\cl$ the rows of $\hat{\mB}$ and $\hat{\mC}$, respectively, the entries of the noiseless measurements $\tilde{\vy}$ can be expressed as
\begin{align*}
\tilde{y}^2_\ell =\tfrac{1}{m} |\<\bl,\vh^\nat\>\<\cl,\vm^\nat\>|^2,\ \ell = 1\ldots m.
\end{align*}
This problem is non-linear in both unknowns; however, it reduces to a bilinear problem in the lifted variables $\vh^\nat\vh^{\nat*}$ and $\vm^\nat\vm^{\nat*}$, taking the form 
\begin{align}\label{eq:lifted-measurements}
\tilde{y}^2_\ell = \tfrac{1}{m}\<\bl\blt,\vh^\nat\vh^{\nat*}\>\<\cl\clt,\vm^\nat\vm^{\nat*}\> = \tfrac{1}{m}\<\bl\blt,\mH^\nat\>\<\cl\clt,\mM^\nat\> ,
\end{align}
where $\mH^\nat = \vh^\nat\vh^{\nat*}$ and $\mM^\nat = \vm^\nat\vm^{\nat*}$. Treating the lifted variables $\mH^\nat$ and $\mM^\nat$ as unknowns makes the measurements bilinear in the unknowns; a structure that will help us formulate an effective convex relaxation. 

In the case of noisy measurements, we will write without loss of generality that
\begin{align}
&\vy = \tfrac{1}{\sqrt{m}}|\hat{\mB}\vh^\nat \odot \hat{\mC}\vm^\nat|\odot(1+\vxi),\label{eq:noisy-measurements} \\
&\qquad\qquad \xi_\ell \geq -1 \ \text{for every} \  \ell = 1\ldots m.\label{eq:noise-conditions}
\end{align}
The noiseless case is given by $\vxi=0$.

\subsection{Novel Convex Relaxation}

The task of recovering $\mH^\nat$ and $\mM^\nat$ from the noiseless measurements $\tilde{\vy}$ in \eqref{eq:noisy-measurements} can be naturally posed as an optimization program 
\begin{align}\label{eq:raw-optimization-program}
&~~~\text{find} ~~~~~~~ \mH, \mM \\
&\text{subject to} ~~ \tfrac{1}{m}\< \bl\blt,\mH\>\< \cl\clt,\mM\> = \tilde{y}^2_\ell , ~ \ell = 1\ldots m.\notag\\
& ~~~\qquad \qquad \text{rank}(\mH) = 1, ~ \text{rank}(\mM) = 1.\notag
\end{align}
Both the measurement and the rank constraints are non-convex.  Further, the immediate convex relaxation of each measurement constraint is trivial, as the convex hull of the set of $(\mH, \mM)$ satisfying $\tilde{y}^2_\ell =  \tfrac{1}{m}\< \bl\blt,\mH\>\< \cl\clt,\mM\>$ is the set of all possible $(\mH, \mM)$.

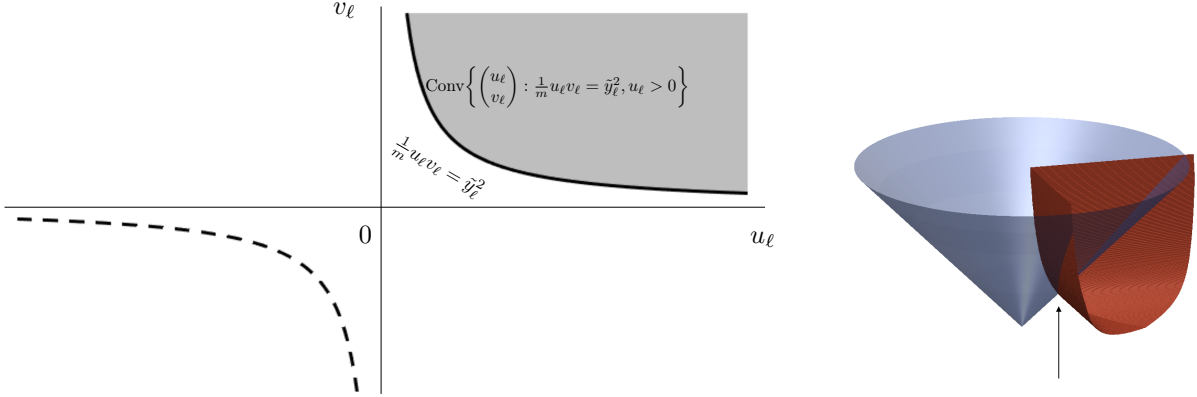
\begin{figure}
	\begin{overpic}[ width=0.57\textwidth,height=0.3\textwidth,tics=10]{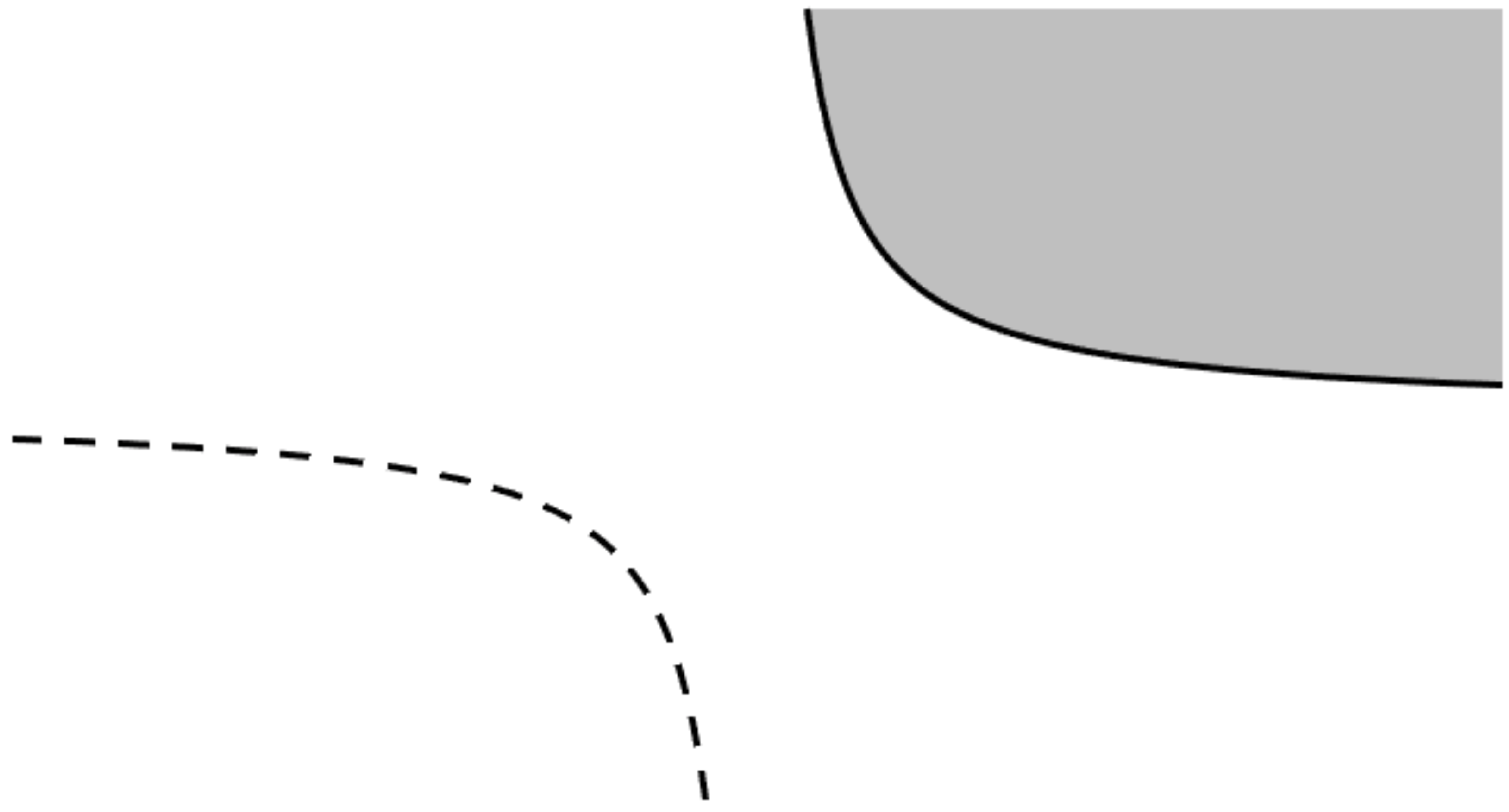}
		\linethickness{0.3pt}
		\put(-1,25.8){\color{black}\line(1,0){103}}
		\put(50,.5){\color{black}\line(0,1){52.5}}
		\put(47,21){$0$}
		\put(43.5,52){$v_\ell$}
		\put(100,21){$u_\ell$}
		\put(51,34){\rotatebox{-30}{\scalebox{.75}{$\tfrac{1}{m}u_\ell v_\ell =\tilde{y}_\ell^2 $}}}
		\put(56,41.5){\rotatebox{0}{\scalebox{.65}{$\mbox{Conv}\!\left\{\begin{pmatrix}u_\ell\\v_\ell\end{pmatrix}: \tfrac{1}{m} u_\ell v_\ell =\tilde{y}_\ell^2, u_\ell>0\right\}$}}}
	\end{overpic}
	\hspace{.7cm}\includegraphics[scale = 0.35]{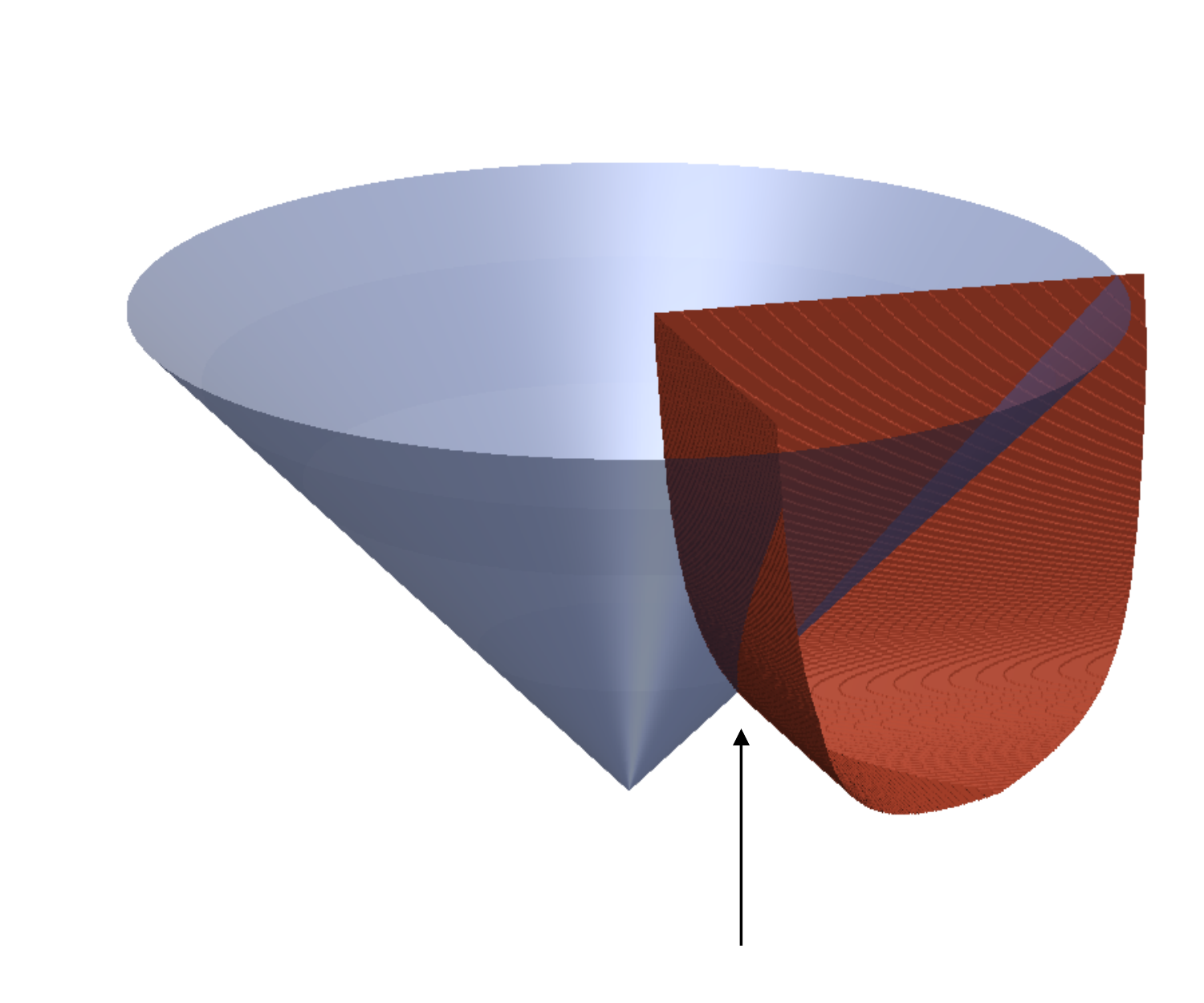}
	\caption{Left: The restriction of the hyperbolic constraint to the first quadrant;  Right: Abstract illustration of the geometry of the convex relaxation. The PSD cone (blue) and the surface of the hyperbolic set (red) formed by two intersecting hyperbolas $(m=2)$. Evidently, there are multiple points on the surface and also in the convex hull of the hyperbolic set that lie on the PSD cone. The minimizer of the optimization program \eqref{eq:convex-optimization-program} picks the one with minimum trace that happens to lie at the intersection of hyperbolic ridge and the PSD cone (pointed out by an arrow).}
	\label{fig:Geometry}
\end{figure}

To derive our convex relaxation, recall that the true $\mH^\nat = \vh^\nat\vh^{\nat*}$, and $\mM^\nat = \vm^\nat\vm^{\nat*}$ are also positive semidefinite (PSD). This means that incorporating the PSD constraint in the optimization program translates into the fact that the variables $u_\ell = \< \bl\blt,\mH\>$ and $v_\ell = \< \cl\clt,\mM\>$ are necessarily non-negative. That is, 
\begin{align*}
\mH \succcurlyeq \mathbf{0}, \ \text{and} \  \mM \succcurlyeq \mathbf{0}  \implies u_\ell \geq 0, \ \text{and} \ v_\ell \geq 0,
\end{align*}
where the implication  follows by the definition of PSD matrices. This observation restricts the hyperbolic constraint set in Figure~\ref{fig:Geometry} to the first quadrant only. For a fixed $\ell$, we propose replacing the non-convex hyperbolic set $\{(u_\ell, v_\ell) \in \R^2\ |\ \tfrac{1}{m} u_\ell v_\ell = \tilde{y}^2_\ell, u_\ell \geq 0 , \ v_\ell \geq 0 \}$ with its convex hull $\{(u_\ell, v_\ell) \in \R^2 \ |\ \tfrac{1}{m} u_\ell v_\ell \geq \tilde{y}^2_\ell, u_\ell \geq 0 , \ v_\ell \geq 0 \}.$  In short, our convex relaxation is possible because the PSD constraint from lifting happens to select a specific branch of the hyperbola given by any particular bilinear measurement, and this single branch has a nontrivial convex hull.

The rest of the convex relaxation is standard, as the rank constraint in \eqref{eq:raw-optimization-program} is then relaxed with a nuclear-norm minimization, which reduces to trace minimization in the PSD case:
\begin{align*}
&\minimize   \ \trace(\mH)+\trace(\mM) \\
&\text{subject to} \ \tfrac{1}{m}\<\bl\blt, \mH\>\<\cl\clt,\mM\> \geq \tilde{y}^2_\ell, \ \ell = 1\ldots m\notag\\
&~~\qquad\qquad  \mH \succcurlyeq \mathbf{0}, \  \mM \succcurlyeq \mathbf{0}\notag.
\end{align*}

In the noiseless or noisy cases, we will study the following program, which only differs in that the noiseless observations are substituted by the possibly noisy ones given from \eqref{eq:noisy-measurements}:
\begin{align}\label{eq:convex-optimization-program}
&\minimize   \ \trace(\mH)+\trace(\mM) \\
&\text{subject to} \ \tfrac{1}{m}\<\bl\blt, \mH\>\<\cl\clt,\mM\> \geq y^2_\ell, \ \ell = 1\ldots m\notag\\
&~~\qquad\qquad  \mH \succcurlyeq \mathbf{0}, \  \mM \succcurlyeq \mathbf{0}\notag.
\end{align}

The convexity of the optimization program above is established in the lemma below. A formal proof of he lemma can be found in Appendix A. 
\begin{lemma}\label{lem:convexity}
	The optimization problem \eqref{eq:convex-optimization-program} is a convex program.
\end{lemma}

\subsection{Main Results}
We consider the case of i.i.d. Gaussian measurements,
\begin{align}\label{eq:BC-Random-Model}
\bl &\sim \text{Normal}(0,  \tfrac{1}{m}\mI_k), \quad 
\cl \sim \text{Normal}(0,  \tfrac{1}{m}\mI_n), \quad \ell = 1, \ldots m.
\end{align}
 We show that with this choice, \eqref{eq:convex-optimization-program} recovers a global scaling $(\alpha \mH^\natural, \alpha^{-1}\mM^\natural)$ of $(\mH^\natural, \mM^\natural)$
The exact value of the unknown scalar multiple $\alpha$ can be characterized for the solution of \eqref{eq:convex-optimization-program}. Observe that the solution $(\widehat{\mH},\widehat{\mM})$ of the convex optimization program in \eqref{eq:convex-optimization-program} obeys $\trace(\widehat{\mH}) = \trace(\widehat{\mM})$. We aim to show that the solution of the optimization program recovers the following scaling $(\tilde{\mH}, \tilde{\mM})$ of the true solution $(\mH^\natural,\mM^\natural)$:
\begin{align}
\tilde{\mH} = \sqrt{\frac{\trace(\mM^\natural)}{\trace(\mH^\natural)}} \mH^\natural, ~ \tilde{\mM} = \sqrt{\frac{\trace(\mH^\natural)}{\trace(\mM^\natural)}} \mM^\natural. \label{eq:scaled-solution}
\end{align}
It is worth noting that $\trace(\tilde{\mH}) = \trace(\tilde{\mM})$,  $\tilde{\mH} = \tilde{\vh}\tilde{\vh}^*$, and $\tilde{\mM} = \tilde{\vm}\tilde{\vm}^*$. 

We show that if $\mB$ and $\mC$ are random, and $m$ is sufficiently large with respect to $k+n$, then the convex program \eqref{eq:convex-optimization-program} stably recovers the true solution $(\mH^\nat, \mM^\nat)$ up to the global bilinear scaling, with high probability. 


\begin{theorem}[Stable Recovery]\label{thm:stable-recovery}
 Given the magnitude only Fourier measurements \eqref{eq:noisy-measurements} of the convolution of two unknown vectors $\vw^\natural$, and $\vx^\natural$ in $\setH^m$ contaminated with additive noise $\vxi$ in $\R^m$. Suppose that $\vw^\natural$, and $\vx^\natural$ are generated as in \eqref{eq:subspace-constraints}, where $\mB$, and $\mC$ are known standard Gaussian matrices as in \eqref{eq:BC-Random-Model}. Assume without loss of generality that noise components $\xi_\ell \geq -1$ for every $\ell = 1,2,3,\ldots, m$. Then for any $t>0$, when $m \geq c (\sqrt{(k+n)}\log m + t)^2$, with probability at least $1-\exp(-\tfrac{1}{2} mt^2)$, the solution $(\widehat{\mH},\widehat{\mM})$ of the convex optimization program in \eqref{eq:convex-optimization-program} obeys
 	\begin{align*}
 	\|\widehat{\mH} - \alpha \mH^\natural\|_F^2+ \|\widehat{\mM} - \alpha^{-1}\mM^\natural\|_F^2 \leq 44^2 \|\vxi\|_\infty (\|\tilde{\mH}\|_F^2 + \|\tilde{\mM}\|_F^2),
 	\end{align*}
where $\alpha = \sqrt{\frac{\trace (\mM^\natural)}{\trace (\mH^\natural)}}$, and $c$ is an absolute constant.
\end{theorem}

As a straightforward special case, for noiseless measurements, solving the proposed convex program would identify the true signals exactly, up to the global bilinear ambiguity, with high probability. 

\begin{corollary}[Exact Recovery]\label{thm:main-theorem}
	Consider the magnitude-only Fourier measurements in \eqref{eq:noiseless-measurements} and a similar setting as Theorem \ref{thm:stable-recovery}. Fixing $t>0$, the convex optimization  in \eqref{eq:convex-optimization-program}  uniquely recovers $(\alpha \mH^\natural,\alpha^{-1}\mM^\natural )$  for $\alpha = \sqrt{\frac{\trace \mM^\natural}{\trace \mH^\natural}}$ 
	with probability at least $1-\exp(-\tfrac{1}{2} mt^2)$ whenever $m \geq c (\sqrt{(k+n)}\log m + t)^2$,
	where $c$ is an absolute constant. 
\end{corollary}
Both Theorem \ref{thm:stable-recovery} and Corollary \ref{thm:main-theorem} establish high probability recovery for phaseless blinear inversion within random subspaces, provided that $m$ on the order of $(k+n)$.  Except for log factors, this sample complexity is optimal.  Proof for the theorem is in the appendix and is based on Rademacher complexity estimates of descent directions objective.

\section{Implementing the Convex Program}\label{sec:Implementation}
A conference paper by the authors \cite{AAH2018Blind} presented an ADMM scheme to address the central convex program \eqref{eq:convex-optimization-program}. One of the main computational challenges with that proposed scheme is that it uses a projection onto the positive semi-definite cone at every ADMM iteration.  Such an operation makes the algorithm prohibitively expensive for large problem sizes. In this section, we consider an alternative ADMM scheme which uses a Burer-Monteiro low-rank factorization \cite{burer2003nonlinear,burer2005local, bhojanapalli2018smoothed} to bypasses the PSD projection and speed up the algorithm convergence\footnote{An implementation of our solver is publicly available at: \url{https://github.com/branchhull/BDPR}}.

To proceed, consider our central convex program
\begin{align}\label{eq:orig}
&\underset{\mX_1,\mX_2}{\text{minimize}}~ \trace(\mX_1) + \trace(\mX_2)\\
&\text{subject to} ~ \left\< \ba{1}\ba{1}^*,\mX_1\right\>\left\< \ba{2}\ba{2}^*,\mX_2\right\> \geq  \delta_\ell\geq 0, \quad \ell = 1 \ldots m\notag \\
& \qquad\qquad ~~~ \mX_1\succcurlyeq \boldsymbol{0},  ~\mX_2 \succcurlyeq \boldsymbol{0}. \notag 
\end{align}
Note that complex-valued positive semidefinite matrices are necessarily Hermitian. For a simpler notation, we define the convex set
\begin{equation}\label{eq:C}
\mathcal{C} = \left\{ \left(\vu,\vv \right)\in\mathbb{R}^m\times\mathbb{R}^m: u_\ell v_\ell\geq \delta_\ell>0,u_\ell\geq 0 \right\}.
\end{equation}
An alternative way of formulating program \eqref{eq:orig} is
\begin{align}\label{eq:cast}
&\underset{\{\mX_j,\vu_j\}_{j=1,2}}{\text{minimize}}~~~   \ind_{\mathcal{C}}(\vu_1,\vu_2)+\sum_{j=1}^2\trace(\mX_j) + \ind_+(\mX_j) \\
&\text{subject to} ~~~~ u_{j,\ell} = \left\< \ba{j}\ba{j}^*,\mX_j\right\>, \notag  ~ \ell = 1\ldots m\notag, ~ j=1,2,
\end{align}
where 
\[\ind_{\mathcal{C}}(\vu,\vv) = \left\{\begin{array}{lc}0 & (\vu,\vv)\in\mathcal{C}\\ +\infty& (\vu,\vv)\notin\mathcal{C}\end{array}  \right., ~~~ \ind_+(\mX) = \left\{\begin{array}{lc}0 & \mX\succeq \boldsymbol{0}\\ +\infty& \mX\nsucceq \boldsymbol{0}\end{array}  \right..
\] 
Defining the dual vectors $\balpha_1, \balpha_2\in \mathbb{R}^m$, the augmented Lagrangian for \eqref{eq:cast} takes the form 
\begin{align}\mathcal{L}\left( \{\mX_j,\vu_j,\balpha_j\}_{j=1,2} \right) &=  \ind_{\mathcal{C}}(\vu_1,\vu_2) + \sum_{j=1}^2\trace(\mX_j) + \ind_+(\mX_j) \notag \\ & ~~~~+ \frac{\rho}{2} \sum_{j=1}^2\sum_{\ell=1}^m \left( u_{j,\ell} - \left\< \ba{j}\ba{j}^*,\mX_j \right\>+ \alpha_{j,\ell}\right)^2.
\end{align}
To set up an ADMM scheme, each variable update at the $k$-th iteration is performed by minimizing $\mathcal{L}$ with respect to that variable while fixing the others. More specifically, using the superscript $(k)$ to denote the iteration, for $j=1,2$ we have the primal updates 
\begin{align}\label{eq:admm1}
&\mX_j^{(k+1)} = \argmin_{\mX_j\succeq\boldsymbol{0}}~~  \trace(\mX_j) + \frac{\rho}{2} \!\sum_{\ell=1}^m\! \left( \left\< \ba{j}\ba{j}^*,\mX_j \right\> - u_{j,\ell}^{(k)} -  \alpha_{j,\ell}^{(k)}\right)^2,\\  &\left(\vu_1^{(k+1)},\vu_2^{(k+1)}\right) = \argmin_{(\vu_1,\vu_2)\ \! \in \ \! \mathcal{C}} ~~ \frac{1}{2}\sum_{j=1}^2\sum_{\ell=1}^m \left( u_{j,\ell} - \left\< \ba{j}\ba{j}^*,\mX_j^{(k+1)} \right\>+ \alpha_{j,\ell}^{(k)}\right)^2,\label{eq:admm2}
\end{align}
along with the dual updates 
\begin{align*}
 \alpha_{j,\ell}^{(k+1)} &= \alpha_{j,\ell}^{(k)} + u_{j,\ell}^{(k+1)} - \left\< \ba{j}\ba{j}^*,\mX_j^{(k+1)} \right\>.  
\end{align*}
In the sequel we outline a computational procedure for each step of the proposed ADMM scheme.  

\subsection{Performing the $\mX$-update}
Central to the ADMM step \eqref{eq:admm1}, in this section we focus on addressing the convex program
\begin{equation}\label{eq:xupdate}
\underset{\mX ~\!\succeq ~\!\boldsymbol{0}}{\text{minimize}}~~ \trace(\mX) + \frac{\rho}{2} \!\sum_{\ell=1}^m\! \left( \va_\ell^*\mX\va_\ell - \theta_\ell\right)^2.
\end{equation}
One of the most successful heuristics to address \eqref{eq:xupdate}, which was brought into attention by \cite{burer2003nonlinear}, is to consider the PSD factorization $\mX = \mV\mV^*$ and to address the non-convex program
\begin{equation}\label{eq:nonconv}
\underset{\mV}{\text{minimize}}~~ \|\mV\|_F^2 + \frac{\rho}{2} \!\sum_{\ell=1}^m\! \left( \left\|\mV^*\va_\ell\right\|_2^2 - \theta_\ell\right)^2.
\end{equation}
For a large class of objectives, there have been theoretical arguments that local minimizers to \eqref{eq:nonconv} can form the global minimizer to \eqref{eq:xupdate}. Specifically, for the objective form \eqref{eq:xupdate}, \cite{bhojanapalli2018smoothed} have recently shown that for almost all objectives of this form, if $\tilde\mV\in\mathbb{R}^{n\times r}$ is a second-order stationary solution to \eqref{eq:nonconv} and $r(r+1)>2m$, then $\tilde\mX = \tilde\mV\tilde\mV^*$ is a global minimizer to \eqref{eq:xupdate} (see Corollary 2 in the aforementioned reference).

Finding solutions to \eqref{eq:nonconv} can be performed via standard optimization toolboxes. In particular, we use quasi-Newton methods with cubic line search as implemented in \cite{schmidt2005minfunc}, which only need the gradient of the objective in \eqref{eq:nonconv}, calculated as
\[2\mV+\rho\sum_{\ell=1}^m  \left( \left\|{\mV}^*\va_\ell\right\|_2^2 - \theta_\ell\right)\va_\ell\va_\ell^*\mV.
\]
It is noteworthy that the gradient calculation only requires a series of matrix-vector multiplications. 

With the proposed computational scheme, to update $\mX$ at each ADMM iteration, another iterative scheme needs to be carried out to solve \eqref{eq:nonconv}. Despite the nested nature of this framework, a very good initialization for $\mV$ at the start of each ADMM update is the optimal $\mV$ from the previous ADMM step. Aside from the factorization technique, such choice of initialization further contributes to fast solutions of \eqref{eq:xupdate}.

\subsection{Performing the $\vu$-update}
The $\vu$-update in \eqref{eq:admm2} is a standard projection problem onto the set $\mathcal{C}$. It is straightforward to see that program
\begin{equation}\min_{(\vu_1,\vu_2)\ \! \in \ \! \mathcal{C}} ~ \frac{1}{2}\sum_{j=1}^2\sum_{\ell=1}^m \left( u_{j,\ell} - \theta_{j,\ell}\right)^2  \label{eq:projhyp}
\end{equation}
decouples into $m$ distinct programs of the form 
\begin{equation}\min_{u_1,u_2} ~ \frac{1}{2}\sum_{j=1}^2\left( u_{j} - \theta_{j}\right)^2 ~~ \mbox{subject to:}~~ u_1u_2\geq \delta> 0, ~ u_1\geq 0. \label{eq:projhypdecoup}
\end{equation}
In the sequel we focus on addressing \eqref{eq:projhypdecoup}, as solving \eqref{eq:projhypdecoup} for each component $\ell$ would deliver the solution to \eqref{eq:projhyp}. We proceed by forming the Lagrangian for the constrained problem \eqref{eq:projhypdecoup}
\[l(u_1,u_2,\mu_1,\mu_2) = \frac{1}{2}\left\|\begin{pmatrix}u_1\\ u_2 \end{pmatrix} -  \begin{pmatrix} \theta_1\\ \theta_2  \end{pmatrix}\right\|_2^2 + \mu_1\left( \delta - u_1 u_2 \right ) - \mu_2u_1.
\]
Along with the primal constraints, the Karush-Kuhn-Tucker optimality conditions are 
\begin{align}
\label{e5}
\frac{\partial l}{\partial u_1} = u_1 - \theta_1 - \mu_1 u_2  - \mu_2 &=0,\\ 
\label{e6}
\frac{\partial l}{\partial u_2} = u_2 - \theta_2 - \mu_1 u_1   &=0,\\ 
\notag 
\mu_1\geq 0, \quad \mu_1\left( \delta - u_1 u_2 \right ) &=0,\\
\notag \mu_2 \geq 0, \quad \mu_2u_1&=0.
\end{align}
We now proceed with the possible cases.

\textbf{Case 1.} $\mu_1=\mu_2=0$:\\
In this case we have $(u_1,u_2)=(\theta_1,\theta_2)$ and this result would only be acceptable when $u_1u_2 \geq \delta$ and $u_1\geq 0$.

\textbf{Case 2.} $\mu_1=0$, $u_1 =0$:\\ 
In this case the first feasibility constraint of \eqref{eq:projhypdecoup} requires that $\delta\leq 0$, which is not a possiblity.

\textbf{Case 3.} $\delta - u_1 u_2 = 0$, $u_1 =0$:\\ 
Similar to the previous case, this cannot happen when $\delta>0$.

\textbf{Case 4.} $\mu_2=0$, $\delta - u_1 u_2  =0$:\\ 
In this case we have $\delta = u_1 u_2$, combining which with \eqref{e6} yields $\delta = (\theta_2 + \mu_1 u_1)u_1$, or
\begin{align}\label{e17}
\mu_1 = \frac{\delta - \theta_2u_1}{u_1^2}.
\end{align}
Similarly, \eqref{e5} yields 
\begin{equation}\label{e19}
u_1 = \theta_1 + \mu_1(\theta_2 + \mu_1 u_1).
\end{equation}
Since the condition $\delta=u_1u_2$ requires that $u_1>0$, $\mu_1$ can be eliminated between \eqref{e17} and \eqref{e19} to generate the following forth order polynomial equation in terms of $u_1$:
\begin{align*}
u_1^4-\theta_1 u_1^3 +\delta\theta_2 u_1- \delta^2=0.
\end{align*}
After solving this fourth order polynomial equation, we pick the real root $u_1$ which obeys
\begin{align}\label{eqconsts}
u_1\geq 0, \qquad \delta - \theta_2u_1\geq 0.
\end{align}
Note that the second inequality in \eqref{eqconsts} warrants nonnegative values for $\mu_1$ thanks to \eqref{e17}. After picking the right root, we can explicitly obtain $\mu_1$ using \eqref{e19} and calculate the $u_2$ using \eqref{e6}. The resulting $(u_1,u_2)$ pair presents the solution to \eqref{eq:projhypdecoup}, and finding such pair for every $\ell$ provides the solution to \eqref{eq:projhyp}. Thanks to the decoupling of the projection step in $\ell$, the $\vu$-update can enjoy a parallel computing framework.

\section{Experiments and Application}
 
\begin{figure}[h]
	\hspace{-.2cm}\begin{overpic}[ height=.24\textwidth,tics=10]{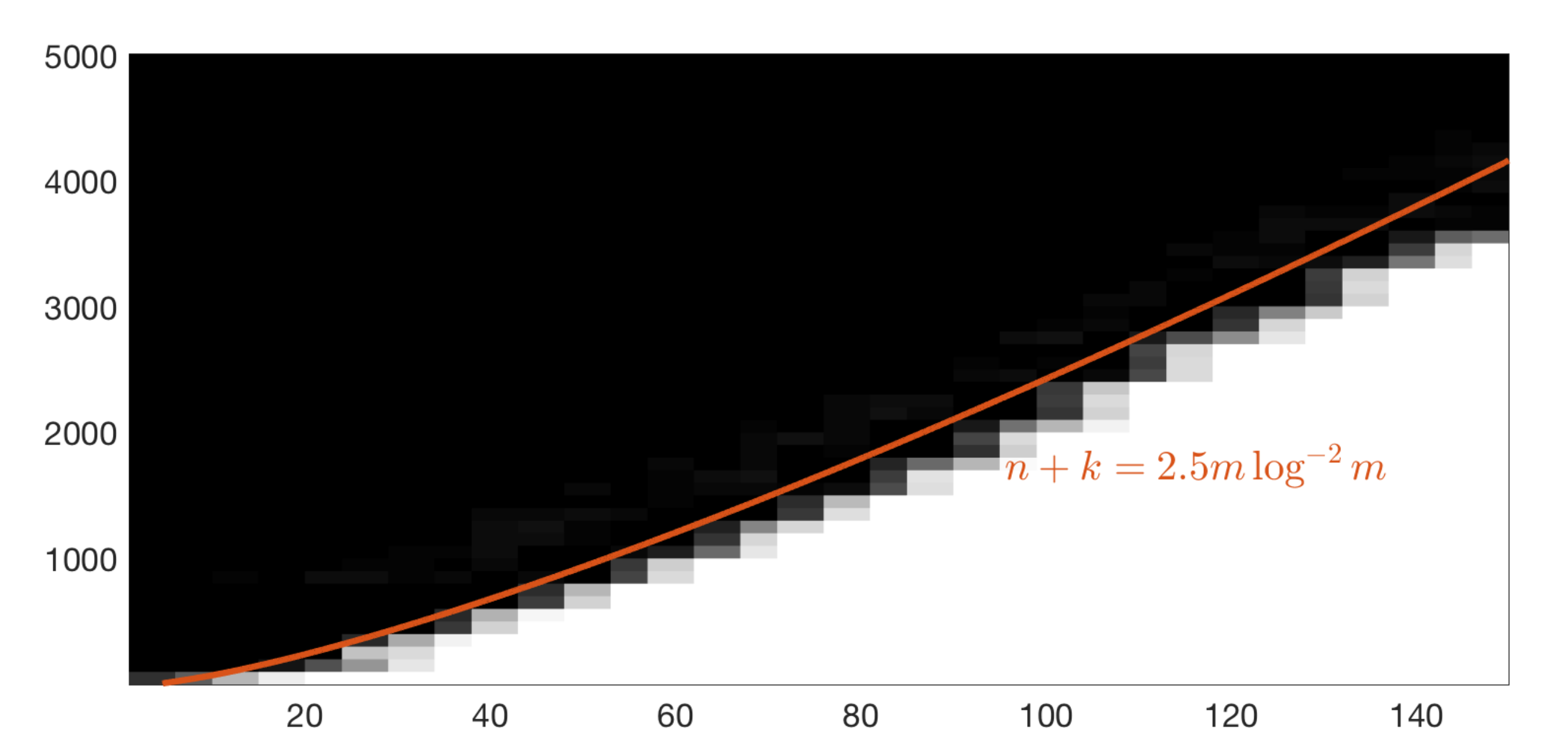}
		\put (49,-2.5) {\scalebox{.85}{$n+k$}}
		\put (-2,25) {\scalebox{.85}{$m$}}
	\end{overpic}
	\begin{overpic}[ height=.24\textwidth,tics=10]{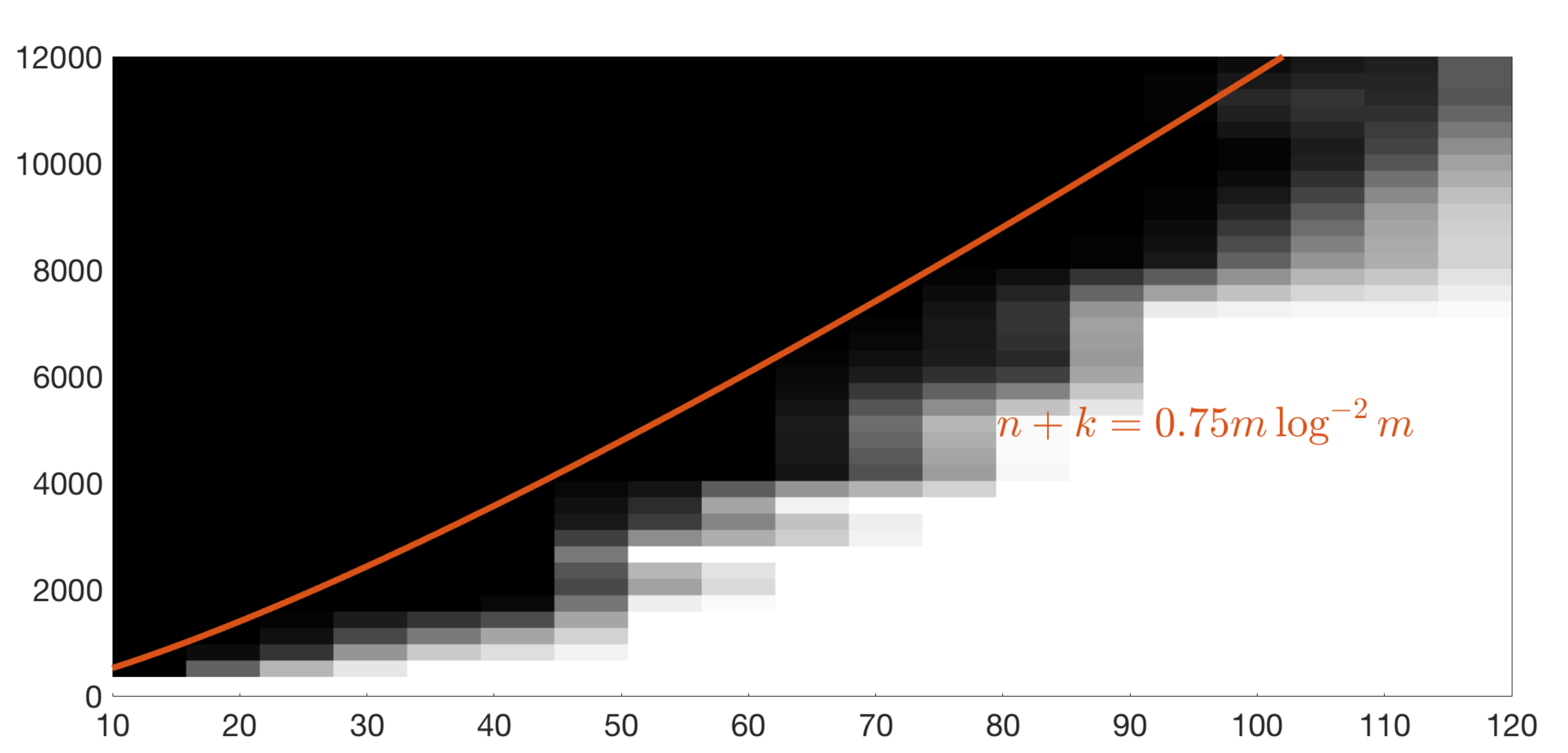}
		\put (49,-2.5) {\scalebox{.85}{$n+k$}}
		\put (-2,25) {\scalebox{.85}{$m$}}
	\end{overpic}
	
	\caption{Phase portraits highlighting the frequency of successful recoveries of the proposed convex program for random and deterministic channel subspaces (see the text for the experiment details).}\label{figphase}
\end{figure}

 We now present numerical experiments that verify the recovery guarantee for bilinear inversion from phaseless Fourier measurements by program \eqref{eq:convex-optimization-program}.  We consider the noiseless case with i.i.d. Gaussian matrices $\mB$ and $\mC$.  In Figure \ref{figphase} we present the phase portrait associated with the proposed convex framework. To obtain the diagram on the left panel, for each fixed value of $m$, we run the algorithm for 100 different combinations of $n$ and $k$, each time using independently generated Gaussian matrices $\mB$ and $\mC$. If the algorithm converges to a sufficiently close neighborhood of the ground-truth solution (a relative error of less than 1\% with respect to the $\ell_2$ norm), we label the experiment as successful. Figure \ref{figphase} shows the collected success frequencies, where solid black corresponds to 100\% success and solid white corresponds to 0\% success.  For an empirically selected constant $c$, the success region almost perfectly stands on the left side of the line $n+k = cm\log^{-2}m$.  The results indicate that the constants in the Theorem are not unreasonably large in practice.

While the analysis in this paper is specifically focused on the Gaussian subspace embeddings for $\vw$ and $\vx$, we additionally consider the case where $\mB$ is deterministic and $\mC$ is Gaussian.  Specifically $\mB$ will be an equispaced sampling of the columns of the identity matrix.   On the right panel of Figure \ref{figphase}, we have plotted the phase diagram for this case of deterministic $\mB$ and random $\mC$. These results hint that the convex framework is applicable to more realistic deterministic subspace models. 

We do not want to give the reader the impression that the present paper solves the problem of blind deconvolutional phase retrieval in practice.  The numerical experiments we perform do indeed show excellent agreement with the theorem in the case of random subspaces.  Such subspaces are unlikely to appear in practice, and typically appropriate subspaces would both be deterministic, including partial Discrete Cosine Transforms or partial Discrete Wavelet Transforms.  Numerical experiments, not shown, indicate that our convex relaxation is less effective for the cases of these doubly deterministic subspaces.  We suspect this is due to the fact that the subspaces for both measurements should be mutually incoherent, in addition to both being incoherent with respect to the Fourier basis given by the measurements.  As with the initial recovery theory for the problems of compressed sensing and phase retrieval, we have studied the random case in order to show information theoretically optimal sample complexity is possible by efficient algorithms.  Based on this work, it is clear that blind deconvolutional phase retrieval is still a very challenging problem in the presence of deterministic matrices, and one for which development of convex or nonconvex methods may provide substantial progress in applications.

\subsection{Related Real-World Applications}
As discussed earlier, the proposed framework addresses a general version of the phase retrieval, where as a result of the light propagation through a medium, the rays are convolved with an unknown kernel. Aside from this general setup, in this section we will point out two specific physical problems, solving which requires simultanuously addressing variants of the phase retrieval and blind deconvolution problems.

\subsubsection{Stylized Application in Visible Light Communications}



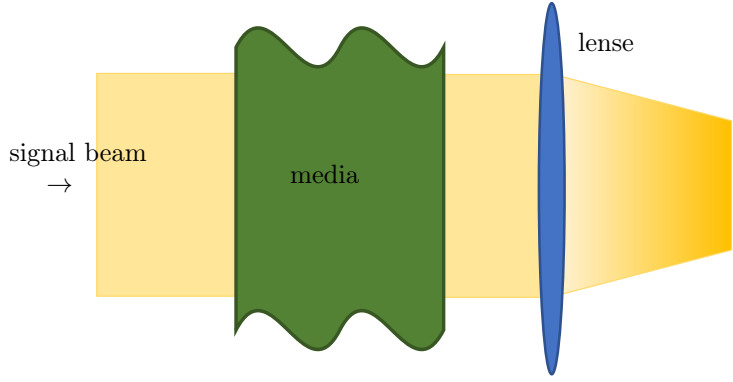
\begin{figure}
	\centering
	\begin{overpic}[ width=0.57\textwidth,height=0.3\textwidth,tics=10]{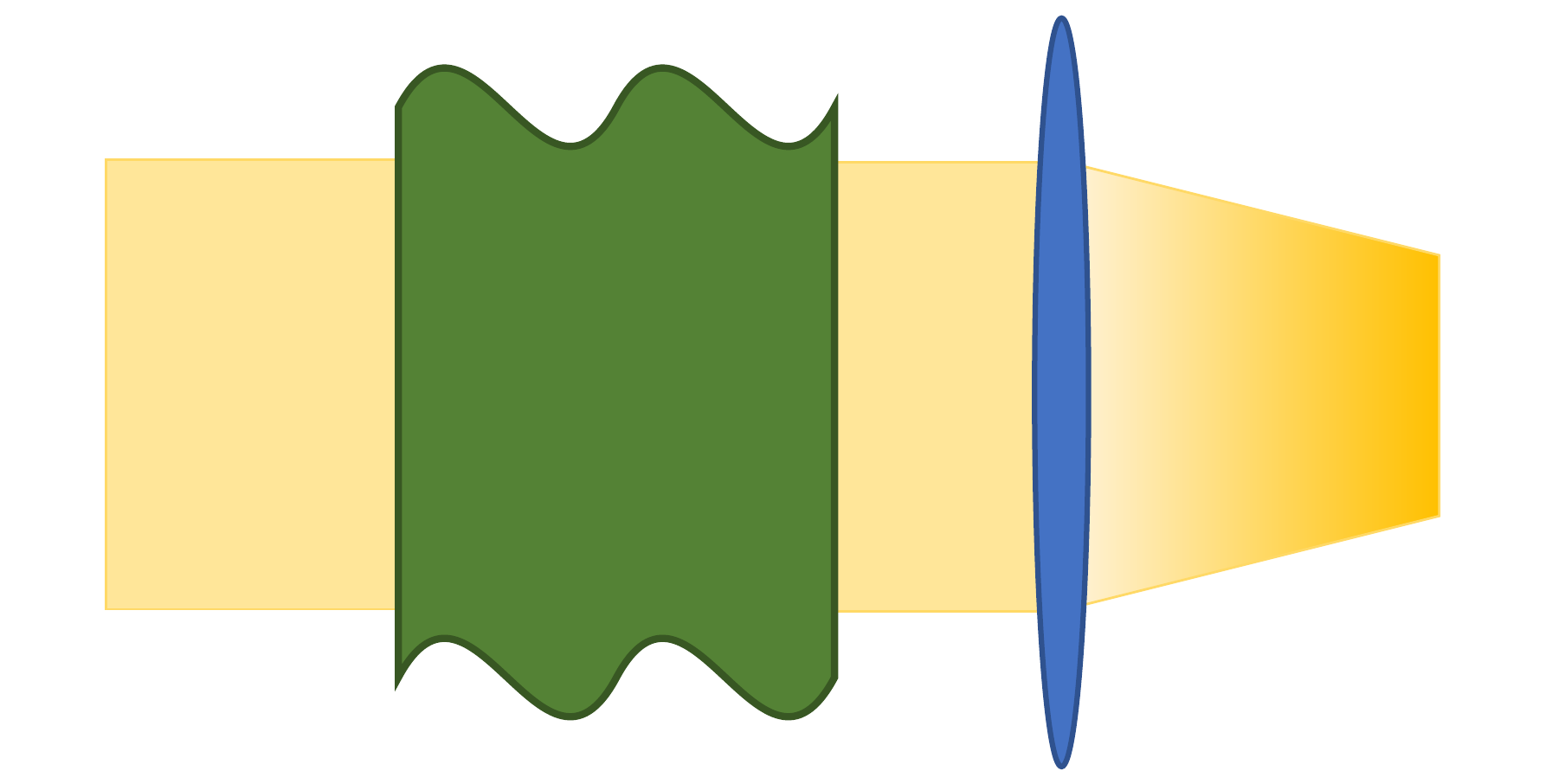}
		\linethickness{0.3pt}
		\put(0,25.85){$\rightarrow$}
		\put(-5,30){signal beam}
		\put(72,45){lense}
		\put(33,27){media}
	\end{overpic}
	\caption{\small Visible light communication optical setup; the media block normally consists of phosphor, filter and a linear polarizer. The lens takes the Fourier transform of the light and one can only measure the intensity only measurements of this transformed light source signal.}
	\label{fig:VLC}
\end{figure}

As discussed in the body of the paper, an important application domain where blind deconvolution from phaseless Fourier measurements arises is the visible light communication (VLC). A stylized VLC setup is shown in Figure \ref{fig:VLC}. A message $\vm \in \R^n$ is to be transmitted using visible light. The message is first coded by multiplying it with a tall coding matrix $\mC \in \R^{m \times n}$ and the resultant information $\vx = \mC\vm$ is modulated on a light wave. The light wave propagates through an unknown media. This propagation can be modeled as a convolution $\vx\circledast \vw$ of the information signal $\vx$ with unknown channel $\vw \in \R^m$. The vector $\vw$ contains channel taps, and frequently in realistic applications has only few significant taps. In this case, one can model 
\[
\vw \approx \mB \vh,
\]
where $\vh \in \R^k$ is a short $(k \ll m)$ vector, and $\mB \in \R^{m \times k}$ in this case is a subset of the columns of an identity matrix. Generally, the multipath channels are well modeled with non-zero taps in top locations of $\vw$. In that case, $\mB$ is exactly known to be the top few columns of the identity matrix. 

In visible light communication, there is always a difficulty associated with measuring phase information in the received light. Figure \ref{fig:VLC} shows a setup, where we measure the phaseless Fourier transform (light through the lens) of this signal. The measurements are therefore
\[
\tilde{\vy} = |\mF (\mB\vh \circledast \mC\vm)|,
\]
and one wants to recover $\vm$, and $\vh$ given the knowledge of $\mB$ and the coding matrix $\mC$.  Since we chose $\mC$ to be random Gaussian, and $\mB$ is the columns of identity. As mentioned at the end of the numerics section that with this subspace model, we obtain similar recovery results as one would have for both $\mB$, and $\mC$ being random Gaussians. The proposed convex program solves this difficult inverse problem and recovers the true solution with these subspace models.  

\subsubsection{Crystallography}
In crystallography, the lattice structural information is carried in the electron density function of the crystal, which may be represented as
\begin{equation}
\rho(\vx) = \sum_{\vy\in S} \rho_c(\vx-\vy).
\end{equation}
Here, $\rho_c(\vx)$ is a compactly supported central motif, and $S$ is a finite, but large compact set of translation vectors. In a sense, the electron density function is the result of convolving the central motif with the indicator of the set $S$. 

Denoting the Fourier transforms of $\rho(\vx)$ and $\rho_c(\vx)$ by $\hat\rho(\boldsymbol{\omega})$ and $\hat\rho_c(\boldsymbol{\omega})$, similar to the other phase retrieval problems, X-ray experiments measure the magnitude of the Fourier transform of $\hat\rho(\boldsymbol{\omega})$, which can be written as 
\[\hat\rho(\boldsymbol{\omega}) = \sum_{\vy\in S}\exp\left( -i2\pi \langle \vx,\boldsymbol{\omega} \rangle\right)\hat\rho_c(\boldsymbol{\omega}).
\]
Identifying the motif $\hat\rho_c(\boldsymbol{\omega})$ and the set $S$, using measurements of the form $\left| \hat\rho_c(\boldsymbol{\omega})\right|^2$ would be a problem which involves simultaneously addressing a phase retrieval and blind deconvolution problem. The reader is referred to 
\cite{elser2017benchmark} and the references therein for more details of the underlying physics and measurement system.

\section{Proof of Theorem \ref{thm:stable-recovery}}
As shown in Appendix \ref{appx:convexity}, the hyperbolic feasible set $\{(\mH,\mM) : \tfrac{1}{m}\<\bl\bl^*,\mH\>\<\cl\cl^*,\mM\> \geq y^2_\ell \}$ is convex in $(\mH,\mM)$, however, the corresponding constraint function\footnote{We will abuse the notation by specifying the same using $(u_\ell,v_\ell)$ as parameters, i.e., $f^{\circ}(u_\ell,v_\ell) = y_\ell^2 -\tfrac{1}{m}u_\ell v_\ell$, where recall that $u_\ell = \<\bl\bl^*,\mH\>$, and $v_\ell = \<\cl\cl^*,\mM\>$} $f^{\circ}_\ell(\mH,\mM) = y^2_\ell - \frac{1}{m} \<\bl\bl^*,\mH\>\<\cl\cl^*,\mM\>$ is a non-convex function of $(\mH,\mM)$. In the analysis later, it is easier to work with convex constraint functions instead; therefore, we replace the function $f^{\circ}_{\ell}(\mH,\mM)$ above with a convex counterpart  whose 0-level-set is the same under the additional constraints that $\mH \succcurlyeq 0$ and $\mM \succcurlyeq 0$:
\begin{align}\label{eq:convex-function-for-constraints}
& f_\ell(\mH,\mM) :=\notag\\
& \gamma_\ell(\mH,\mM)\left( \sqrt{4y^2_\ell +\frac{1}{m} (\<\bl\bl^*,\mH\>-\<\cl\cl^*,\mM\>)^2 } - \frac{1}{\sqrt{m}}(\<\bl\bl^*,\mH\> +\<\cl\cl^*,\mM\>)\right),
\end{align}
where 
\begin{align}\label{eq:gamma-def}
\gamma_\ell(\mH,\mM) := \begin{cases}
\min \left\{ 1, \frac{ \<\bl\bl^*,\tilde{\mH}\> + \<\cl\cl^*,\tilde{\mM}\>}{2}\right\}, & \ f_\ell(\mH,\mM) \leq 0 \\
\frac{ \<\bl\bl^*,\tilde{\mH}\> + \<\cl\cl^*,\tilde{\mM}\>}{2}, & \text{otherwise}.
\end{cases}
\end{align}
is a scalar chosen to normalize the gradients computed below. Recall that $y^2_\ell = \tilde{y}_\ell^2(1+\xi_\ell)$. It is now easy to check that feasible sets drawn by the convex and non-convex functions are equal under the additional constraint of $u_\ell = \<\bl\blt,\mH\> \geq 0$, and $v_\ell = \<\cl\clt,\mM\> \geq 0$ for every $\mH$, and $\mM$, i.e., $\mH\succcurlyeq \mathbf{0}$, and $\mM \succcurlyeq\mathbf{0}$, respectively. Mathematically, 
\begin{align*}
&  \left\{ (u_\ell,v_\ell) \ | \ f^{\circ}_\ell(u_\ell,v_\ell) := y^2_{\ell} - \tfrac{1}{m}u_\ell v_\ell \leq 0 , u_\ell \geq 0, \ v_\ell \geq 0\right\} = \\
& \left\{(u_\ell,v_\ell) \ | \ f_\ell(u_\ell, v_\ell) := \gamma_\ell\left(\sqrt{4y^2_\ell+\tfrac{1}{m}(u_\ell-v_\ell)^2} - \tfrac{1}{\sqrt{m}}(u_\ell+v_\ell)\right) \leq 0 \right\},
\end{align*}
for any $\gamma_\ell > 0$. 
Note that $f_\ell (u_\ell,v_\ell) \leq 0$ automatically constrains $u_\ell \geq 0$ and $v_\ell \geq 0$.
It is easy to check that $f_\ell (u_\ell,v_\ell)$ is a convex function. 
Since $\mH \succcurlyeq 0$ and $\mM \succcurlyeq \mathbf{0}$ imply that $u_\ell \geq 0$,  and $v_\ell \geq 0$, respectively, and since $\gamma_\ell(\mH,\mM) \geq 0$, we can write the above conclusion in the matrix space as
\begin{align*}
&  \bigg\{ (\mH,\mM) \in \setH^{k \times k} \times \setH^{n \times n} ~ \big|~ y^2_{\ell} \leq \tfrac{1}{m}\<\bl\bl^*,\mH\>\<\cl\cl^*,\mM\>, 
\mH \succcurlyeq 0, \ \mM \succcurlyeq 0\bigg\} = \\
& \qquad \bigg\{(\mH,\mM) \in \setH^{k \times k} \times \setH^{n \times n} ~ \big|~ f_\ell(\mH,\mM) \leq 0, \mH \succcurlyeq 0, \ \mM \succcurlyeq 0\bigg\}.
\end{align*}
In the sequel, we will refer to 
\begin{align*}
&\tilde{f}_\ell(\mH,\mM) := \\
& \qquad \gamma_\ell(\mH,\mM)\left( \sqrt{4\tilde{y}^2_\ell +\tfrac{1}{m} (\<\bl\bl^*,\mH\>-\<\cl\cl^*,\mM\>)^2 } - \tfrac{1}{\sqrt{m}}(\<\bl\bl^*,\mH\> +\<\cl\cl^*,\mM\>)\right), 
\end{align*}
which is same as $f_\ell(\mH,\mM)$ except the measurements $y_\ell^2$ is now replaced by corresponding noiseless measurements $\tilde{y}_\ell^2$. Define a convex indicator function for the positive semidefinite cone: 
\begin{align*}
\iota(\mH,\mM) := \begin{cases}
0, & \mH \succcurlyeq 0 \text{ and } \mM \succcurlyeq 0\\
+ \infty, & \text{otherwise.} 
\end{cases}
\end{align*}
Introduce the convex regularizer
\begin{align*}
 \setJ(\mH,\mM) = \trace(\mH) + \trace(\mM)+ \iota(\mH,\mM). 
\end{align*}
For analysis purposes, we will work with the following optimization program 
\begin{align}\label{eq:modified-convex-program}
&\minimize \ \setJ(\mH,\mM) \\
&\text{subject to} \ f_\ell(\mH,\mM) \leq 0, \ \ell = 1 \ldots m,\notag
\end{align}
where  $f_\ell(\mH,\mM)$ is given by \eqref{eq:convex-function-for-constraints}.The optimization program \eqref{eq:modified-convex-program} is equivalent to \eqref{eq:convex-optimization-program} as the objective and constraint set remain unchanged. In the analysis later, we will also need the subdifferential $\nabla \tilde{f}_\ell$, evaluated at $(\tilde{\mH},\tilde{\mM})$, which are given by \eqref{eq:scaled-solution}.
One can verify that 
\begin{align}\label{eq:gradient-ftilde}
-\frac{1}{\sqrt{m}}(\<\cl\cl^*,\tilde{\mM}\>\bl\bl^*, \<\bl\bl^*,\tilde{\mH}\> \cl\cl^*) \in \nabla \tilde{f}_\ell(\tilde{\mH},\tilde{\mM}). 
\end{align}
To see this, refer to a brief derivation below
\begin{align*}
\frac{\partial \tilde{f}_\ell }{\partial \mH}(\tilde{\mH},\tilde{\mM}) &= \gamma_\ell (\tilde{\mH},\tilde{\mM}) \Bigg(  \frac{\frac{1}{m}(\<\bl\bl^*,\tilde{\mH}\>-\<\cl\cl^*,\tilde{\mM}\>)\bl\bl^*}{\sqrt{4\tilde{y}^2_\ell + \frac{1}{m}(\<\bl\bl^*,\tilde{\mH}\>-\<\cl\cl^*,\tilde{\mM}\>)^2 }}  - \frac{1}{\sqrt{m}}\bl\bl^*\Bigg)\\
&= -\frac{1}{\sqrt{m}}\<\cl\cl^*,\tilde{\mM}\>\bl\bl^*,
\end{align*}
where the last equality follows by using $\tilde{y}^2_\ell = \frac{1}{m}\<\bl\bl^*,\tilde{\mH}\>\<\cl\cl^*,\tilde{\mM}\>$.

We now build some preliminaries required to characterize the set of descent directions for the objective function of the optimization program \eqref{eq:modified-convex-program}.  Let $\Th$, and $\Tm$ be the set of symmetric matrices of the form 
\begin{align*}
\Th := \{ \mX = \tilde{\vh}\vz^* + \vz\tilde{\vh}^*\}, ~ \Tm := \{ \mX = \tilde{\vm}\vz^* + \vz\tilde{\vm}^*\},
\end{align*}
and denote the orthogonal complements by $\Thp$, and $\Tmp$, respectively. Note that $\mX \in \Th^\perp$ iff both the row and column spaces of $\mX$ are perpendicular to $\tilde{\vh}$. $\PTh$ denotes the orthogonal projection onto the set $\Th$, and a matrix $\mX$ of appropriate dimensions can be projected into $\Th$ as
\[
\PTh(\mX) : = \tfrac{\tilde{\vh}\tilde{\vh}^* }{\|\tilde{\vh}\|_2^2} \mX+ \mX \tfrac{\tilde{\vh}\tilde{\vh}^* }{\|\tilde{\vh}\|_2^2}  - \tfrac{\tilde{\vh}\tilde{\vh}^* }{\|\tilde{\vh}\|_2^2}\mX\tfrac{\tilde{\vh}\tilde{\vh}^* }{\|\tilde{\vh}\|_2^2}.
\]
Similarly, define the projection operator $\PTm$. The projection onto orthogonal complements are then simply $\setP_{\Thp} : = \setI - \PTh$, and  $\setP_{\Tmp}: = \setI - \PTm$, where $\setI$ is the identity operator. We use $\mX_{\Th}$ as a shorthand for $\PTh(\mX)$. The subgradient $\partial \setJ(\mH,\mM)$ of the objective $\setJ(\mH,\mM)$ at the proposed solution $(\tilde{\mH}, \tilde{\mM})$ is 
\begin{align}\label{eq:objective-subgradient}
\partial \setJ (\tilde{\mH},\tilde{\mM}) &=\Big\{\mG \in \setH^{k \times k} \times \setH^{n \times n}~\big|~\mG = \notag\\
&\left(\tfrac{\tilde{\mH}}{\|\tilde{\mH}\|_F},\tfrac{\tilde{\mM}}{\|\tilde{\mM}\|_F} \right)+ (\mW_{1,\Thp}, \mW_{2,\Tmp}), ~ \lambda_{\max}(\mW_{1,\Thp}, \mW_{2,\Tmp}) \leq 1\Big\},
\end{align}
for details; see, Section 8.6 in \cite{tropp2015convex}, and references therein. 

Given the measurements \eqref{eq:noisy-measurements}, one can only identify the true solution $(\mH^\natural,\mM^\natural)$ up to the blinear scaling ambiguity. To formalize this, begin by defining a set 
\begin{align}\label{eq:setN}
\setN: = \{ (\mH,\mM) \in \setH^{k \times k} \times \setH^{n \times n}~\big| ~ (\mH,\mM) = \beta (-\tilde{\mH},\tilde{\mM}),\beta \in \R\}, 
\end{align}
and denote by $(\tilde{\mH},\tilde{\mM}) \oplus \setN$ a set $\setN$ shifted by a point $(\tilde{\mH},\tilde{\mM})$. Mathematically, 
\begin{align}\label{eq:oplusN}
(\tilde{\mH},\tilde{\mM}) \oplus \setN = \{ (\mH,\mM) \in \setH^{k \times k} \times \setH^{n \times n}~\big| ~(\mH,\mM) = ((1-\beta)\tilde{\mH},(1+\beta)\tilde{\mM}), \beta \in \R\}. 
\end{align}
We will refer to this set as the linearized global scaling of $(\tilde{\mH},\tilde{\mM})$. 

The main argument of stable recovery in the noisy case is summarized as follows:  Let
 \[
 \setN_{\perp}: = \{(\mH,\mM) \in \setH^{k \times k} \times \setH^{n \times n}~ \big| ~ \<\mH,-\tilde{\mH}\> + \<\mM,\tilde{\mM}\> = 0 \}
 \]
 be the orthogonal complement of the subspace $\setN$.  The \textbf{first step} consists of showing that any feasible perturbation $(\dH,\dM) \in \setN_\perp$ about the linearized scaling ambiguity $(\tilde{\mH},\tilde{\mM}) \oplus \setN$ cannot be too large.  This only shows that a large perturbation in the $\setN_\perp$ direction is not allowed, however, the movement away from the ground truth $(\tilde{\mH},\tilde{\mM})$ along the line $(\tilde{\mH},\tilde{\mM}) \oplus \setN$ can still be arbitraritly large. In the \textbf{second step}, we note that the straight line $(\tilde{\mH},\tilde{\mM}) \oplus \setN$ touches (in the noiseless case) the hyperbolic feasible set at $(\tilde{\mH},\tilde{\mM})$, and diverges away from it as we deviate away (large $\beta$) from the point $(\tilde{\mH},\tilde{\mM})$ along the line $(\tilde{\mH},\tilde{\mM}) \oplus \setN$. However, moving too far away along the line  $(\tilde{\mH},\tilde{\mM}) \oplus \setN$ would make it impossible to jump back into the hyperbolic feasible region while not exceeding the allowed leap length in the  $\setN_\perp$ direction prescribed in the first step. This allows us to control deviations from $(\tilde{\mH},\tilde{\mM})$ along $(\tilde{\mH},\tilde{\mM}) \oplus \setN$ as well. Combining the limited allowed deviations both in $\setN$, and $\setN_\perp$ from the ground truth $(\tilde{\mH},\tilde{\mM})$ enables us to show that program recovers a solution $(\widehat{\mH},\widehat{\mM})$ in the neighborhood of $(\tilde{\mH},\tilde{\mM})$. In the noiseless case, the same argument leads to an exact recovery result.

We now formally proceed with the proof argument. The set $\setQ$ of descent directions only in $\setN_\perp$ of the objective function in \eqref{eq:modified-convex-program} is characterized as follows 
\begin{align}\label{eq:setQ}
&\big\{ (\dH,\dM) \in \setN_{\perp} ~\big| ~ \big<(\mG, (\dH,\dM)\big\> \leq 0,  \forall \mG \in \partial \setJ(\tilde{\mH},\tilde{\mM})\big\}\subseteq \notag\\
& \Bigg\{ (\dH,\dM)\in \setN_{\perp} ~\bigg| ~ \Big<\bigg(\tfrac{\tilde{\mH}}{\|\tilde{\mH}\|_F},\tfrac{\tilde{\mM}}{\|\tilde{\mM}\|_F}\bigg),(\dH_{\Th},\dM_{\Tm})\Big> + \trace(\dH_{\Thp}, \dM_{\Tmp} ) \leq 0\Bigg\}\red{\subseteq}  \notag\\
&\big\{ (\dH,\dM)\in \setN_{\perp} ~\big| ~  \trace(\dH_{\Thp}, \dM_{\Tmp} ) \leq  \sqrt{2}\| (\dH_{\Th}, \dM_{\Tm} )\|_F \big\} =: \setQ,
\end{align}
where the first set containment follows by using $\< (\mW_{1,\Thp},\mW_{2,\Tmp}), (\dH_{\Thp},\dM_{\Tmp}) \> \leq \trace(\dH_{\Thp},\dM_{\Tmp})$, which follows from $\lambda_{\max}(\mW_{1,\Thp},\mW_{2,\Tmp}) \leq 1$,  and $(\dH_{\Thp},\dM_{\Tmp}) \succcurlyeq \mathbf{0}$ as any feasible perturbation must obey $(\tilde{\mH}+\dH, \tilde{\mM}+\dM) \succcurlyeq \mathbf{0}$. Last containment simply uses Cauchy-Schwartz inequality, and the fact that
\begin{align*}
\bigg\|\bigg(\tfrac{\tilde{\mH}}{\|\tilde{\mH}\|_F},\tfrac{\tilde{\mM}}{\|\tilde{\mM}\|_F}\bigg)\bigg\|_F = \sqrt{2}. 
\end{align*}

We quantify the "width" of the set of descent directions $\setQ$ through a Rademacher complexity, and a probability that the gradients $\nabla \tilde{f}_\ell$ in \eqref{eq:gradient-ftilde} of the constraint functions of \eqref{eq:modified-convex-program} lie in a certain half space. This enables us to build an argument using the small ball method \cite{koltchinskii2015bounding,mendelson2014learning} that it is unlikely to have points that meet the constraints in \eqref{eq:modified-convex-program} and still be in $\setQ$. Before moving forward, we introduce the above mentioned Rademacher complexity and probability term.   

For a set $\setQ \subset (\setH^{k \times k}, \setH^{n \times n})$, the Rademacher complexity of the gradients $\nabla \tilde{f}_\ell$ in \eqref{eq:gradient-ftilde} is defined as 
\begin{align}\label{eq:Rademacher-Complexity}
\mathfrak{C}(\setQ) := \E \sup_{(\mH,\mM) \in \setQ} \sum_{\ell=1}^m \varepsilon_{\ell}\left\< \nabla \tilde{f}_\ell, \tfrac{(\mH,\mM)}{\|(\mH,\mM)\|_F}\right\>,
\end{align}
where $\varepsilon_\ell, ~\ell = 1\ldots m$ are iid Rademacher random variables independent of everything else in the expression. For a convex set $\setQ$, $\mathfrak{C}(\setQ)$ is a measure of the width of $\setQ$ around origin in terms of the gradients $\nabla \tilde{f}_\ell, ~ \ell = 1\ldots m$. For example, random choice of gradient might yield little overlap with a structured set $\setQ$ leading to a smaller complexity $\mathfrak{\setQ}$. 

Our result also depends on a probability $p_{\tau}(\setQ)$ and a positive parameter $\tau$ defined as 
\begin{align}\label{eq:p_tau}
p_{\tau}(\setQ) := \inf_{(\mH,\mM) \in \setQ} \mathbb{P}\Big( \<\nabla \tilde{f} , (\mH, \mM)\> \geq \frac{\tau}{\sqrt{m}} \|(\mH,\mM)\|_F \Big).
\end{align}
The probability $p_{\tau}(\setQ)$ quantifies visibility of the set $\setQ$ through the gradient vectors\footnote{We drop the subscript $\ell$ here as $\nabla \tilde{f}_\ell, ~ \ell = 1\ldots m$ are identically distributed} $\nabla \tilde{f}$. A small value of $\tau$ and $p_{\tau}(\setQ)$ means that the set $\setQ$ mainly remains invisible through the lenses of $\nabla \tilde{f}_\ell, \ell = 1,2,3, \ldots, m$. This can be appreciated just by noting that $p_\tau(\setQ)$ depends on the correlation of the elements of $\setQ$ with the gradient vectors $\nabla \tilde{f}_\ell$. 

The following lemma shows that the minimizer of  \eqref{eq:modified-convex-program} almost always resides in the desired set $(\tilde{\mH},\tilde{\mM})\oplus \setN$ for a sufficiently large $m$ quantified interms of $\mathfrak{C}(\setQ)$, $p_{\tau}(\setQ)$, and $\tau$. 
\begin{lemma}\label{lem:Sample-Complexity}
	Given the noisy measurements \eqref{eq:noisy-measurements}, where the addtitive noise $\vxi$ obeys \eqref{eq:noise-conditions}. For signal recovery, consider the optimization program in \eqref{eq:modified-convex-program}, and let $\setQ$, characterized in \eqref{eq:setQ}, be the set of descent directions for which $\mathfrak{C}(\setQ)$, and $p_{\tau}(\setQ)$ can be determined using \eqref{eq:Rademacher-Complexity} and \eqref{eq:p_tau}. Choose 
	\begin{align*}
	m \geq \left(\frac{2\mathfrak{C}(\setQ)+t\tau}{\tau p_{\tau(\setQ)}}\right)^2
	\end{align*}
	for any $t >0$. Then the minimizer $(\widehat{\mH}, \widehat{\mM})$ of \eqref{eq:modified-convex-program} satisfies
	\begin{align*}
	\|(\widehat{\mH},\widehat{\mM})-(\tilde{\mH},\tilde{\mM})\|_F^2 \leq 44^2 \|\vxi\|_\infty \left(\|\tilde{\mH}\|_F^2 + \|\tilde{\mM}\|_F^2\right)
	\end{align*}
	with probability at least $1-\mathrm{e}^{-2mt^2}$. 
\end{lemma}
Proof of this lemma is based on small ball method developed in \cite{koltchinskii2015bounding,mendelson2014learning} and further studied in  \cite{lecue2018regularization,lecue2017regularization}. The proof is repeated using the argument in \cite{bahmani2017anchored}, and is provided in the supplementary material for completeness.

Lemma \ref{lem:Sample-Complexity} proves that under the choice of $m$ outlined in Lemma \ref{lem:Sample-Complexity}, the optimization program  \eqref{eq:convex-optimization-program}  recovers $(\tilde{\mH},\tilde{\mM})$ exactly in the noiseless case $\vxi = \boldsymbol{0}$, and stably in the noisy case. The last missing piece in the proof of Theorem \ref{thm:stable-recovery} is to quantify the Rademacher complexity $\mathfrak{C}(\setQ)$, and $p_\tau(\setQ)$ for the $\setQ$ appearing in the measurement bound. 

\subsection{Rademacher Complexity}
We begin with evaluation of the complexity $\mathfrak{C}(\setQ)$
\begin{align*}
\mathfrak{C}(\setQ) &:= \E \sup_{(\dH,\dM) \in \setQ} ~ \sum_{\ell=1}^m \varepsilon_\ell \Big\< \nabla \tilde{f}_\ell, \tfrac{(\dH,\dM)}{\|(\dH,\dM)\|_F}\Big\>
\end{align*}
Splitting $(\dH,\dM)$ between $(\Th,\Tm)$, and $(\Th^\perp,\Tm^\perp)$, and using Holder's inequalities, we obtain
\begin{align*}
&\mathfrak{C}(\setQ)  \leq\\
&  \E \Big\| \tfrac{1}{\sqrt{m}}\sum_{\ell=1}^m\varepsilon_\ell ( \<\cl\cl^*,\tilde{\mM}\> \PTh(\bl\blt),\<\bl\bl^*,\tilde{\mH}\> \PTm(\cl\clt))
\Big\|_F \cdot \sup_{(\dH,\dM) \in \setQ}~ \left\| \tfrac{(\dH_{\Th},\dM_{\Tm})}{\|(\dH,\dM)\|_F}\right\|_F\\
&+ \E \Big\| \frac{1}{\sqrt{m}}\sum_{\ell=1}^m\varepsilon_\ell ( \<\cl\cl^*,\tilde{\mM}\>\bl\blt,\<\bl\bl^*,\tilde{\mH}\>\cl\clt)
\Big\|\cdot \sup_{(\dH,\dM) \in \setQ}~ \left\|\tfrac{ \big(\dH_{\Thp}, \dM_{\Tmp }\big)}{\| (\dH, \dM )\|_F}\right\|_*.
\end{align*}
Recall that $(\dH_{\Thp}, \dM_{\Tmp })\succcurlyeq \mathbf{0},$ and, therefore, $\trace(\dH_{\Thp}, \dM_{\Tmp }) = \|(\dH_{\Thp}, \dM_{\Tmp })\|_*$. On the set $\setQ$, defined in \eqref{eq:setQ}, we have 
\begin{align*}
\frac{\trace(\dH_{\Thp}, \dM_{\Tmp })}{\| (\dH, \dM )\|_F}\leq \sqrt{2}\Bigg\| \frac{\big(\dH_{\Th},\dM_{\Tm}\big)}{\|(\dH,\dM)\|_F}\Bigg\|_F \leq \sqrt{2}.
\end{align*}
Using Jensen's inequality, the first expectation simply becomes
\begin{align*}
&\E\Big\| \frac{1}{\sqrt{m}}\sum_{\ell=1}^m\varepsilon_\ell \big( \<\cl\cl^*,\tilde{\mM}\> \PTh(\bl\blt),\<\bl\bl^*,\tilde{\mH}\> \PTm(\cl\clt)\big)
\Big\|_F \\
&\leq \sqrt{ \frac{1}{m} \E\Big\|\sum_{\ell=1}^m \varepsilon_\ell \big( \<\cl\cl^*,\tilde{\mM}\> \PTh(\bl\blt),\<\bl\bl^*,\tilde{\mH}\> \PTm(\cl\clt)\big)\Big\|_F^2}\\
& = \sqrt{ \frac{1}{m} \sum_{\ell=1}^m \E \Big(\<\cl\cl^*,\tilde{\mM}\>\| \PTh(\bl\blt)\|_F^2+\|\<\bl\bl^*,\tilde{\mH}\> \PTm(\cl\clt)\|_F^2\Big)},
\end{align*}
where the last equality follows by going through with the expectation over $\varepsilon_\ell$'s. Recall from the definition of the projection operator that $\PTh(\bl\blt) : = \tfrac{\tilde{\vh}\tilde{\vh}^* }{\|\tilde{\vh}\|_2^2}\bl\blt + \bl\blt \tfrac{\tilde{\vh}\tilde{\vh}^* }{\|\tilde{\vh}\|_2^2}  - \tfrac{\tilde{\vh}\tilde{\vh}^* }{\|\tilde{\vh}\|_2^2}\bl\blt\tfrac{\tilde{\vh}\tilde{\vh}^* }{\|\tilde{\vh}\|_2^2}$. It can be easily verified that 
$\|\PTh(\bl\blt)\|_F^2 = 2\tfrac{|\blt\tilde{\vh}|^2}{\|\tilde{\vh}\|_2^2} \|\bl\|_2^2 - \tfrac{|\blt\tilde{\vh}|^4}{\|\tilde{\vh}\|_2^4},$
and, therefore, 
\begin{align*}
\E \|(\<\cl\cl^*,\tilde{\mM}\> \PTh(\bl\blt)\|_F^2 &\leq \E |\clt\tilde{\vm}|_2^4 \cdot \E \left(2\tfrac{|\blt\tilde{\vh}|^2}{\|\tilde{\vh}\|_2^2} \|\bl\|_2^2 - \tfrac{|\blt\tilde{\vh}|^4}{\|\tilde{\vh}\|_2^4}\right) \leq 3\|\tilde{\vm}\|_2^4\left(6k-3\right),
\end{align*}
where we used a simple calculation involving fourth moments of Gaussians $\E |\blt\tilde{\vh}|^2 \|\bl\|_2^2 = 3 k \|\tilde{\vh}\|_2^2$. In an exactly similar manner, we can also show that $\|(\<\bl\bl^*,\tilde{\mH}\> \PTm(\cl\clt)\|_F^2 \leq 3\|\tilde{\vh}\|_2^4 (6n-3) $. Putting these together gives us 
\begin{align*}
\E\Big\| \frac{1}{\sqrt{m}}\sum_{\ell=1}^m\varepsilon_\ell \big( \<\cl\cl^*,\tilde{\mM}\> \PTh(\bl\blt),\<\bl\bl^*,\tilde{\mH}\> \PTm(\cl\clt)\big)
\Big\|_F \leq 5\max(\|\tilde{\vh}\|_2^2, \|\tilde{\vm}\|_2^2)\sqrt{k+n}. 
\end{align*}
Moreover, 
\begin{align*}
& \E \Big\| \tfrac{1}{\sqrt{m}} \sum_{\ell=1}^m \varepsilon_\ell \big( \<\cl\cl^*,\tilde{\mM}\>\bl\blt,\<\bl\bl^*,\tilde{\mH}\> \cl\clt\big)
\Big\| \leq \\
& \qquad\qquad \E\max_{\ell}(\<\bl\bl^*,\tilde{\mH}\>,\<\cl\cl^*,\tilde{\mM}\>) \cdot \E\Big\| \tfrac{1}{\sqrt{m}} \sum_{\ell=1}^m \varepsilon_\ell   (\bl\blt,\cl\clt)\Big\|.
\end{align*}
Standard net arguments; see, for example, Sec. 5.4.1 of  \cite{eldar2012compressed} show that 
\begin{align*}
\PP \left(\Big\| \tfrac{1}{\sqrt{m}} \sum_{\ell=1}^m \varepsilon_\ell  (\bl\blt,\cl\clt)\Big\| \geq c \sqrt{k+n}\right) \leq \mathrm{e}^{-c m}, ~ \text{provided that} ~ m \geq c (k+n).
\end{align*}
This directly implies that $\E \Big\| \tfrac{1}{\sqrt{m}} \sum_{\ell=1}^m \varepsilon_\ell  (\bl\blt,\cl\clt)\Big\| \leq c\sqrt{k+n}.$ The random variables $u_\ell$ and $v_\ell$ being sub-exponential have Orlicz-1 norms bounded by $c\max(\|\tilde{\vh}\|_2^2,\|\tilde{\vm}\|_2^2)$. Using standard results, such as Lemma 3 in \cite{van2013bernstein}, we then have $\E \max_{\ell} (u_\ell,v_\ell) \leq c\log m.$ Putting these together yields
\begin{align}
\E \Big\| \tfrac{1}{\sqrt{m}} \sum_{\ell=1}^m \varepsilon_\ell \big( \<\cl\cl^*,\tilde{\mM}\>\bl\blt,\<\bl\bl^*,\tilde{\mH}\> \cl\clt\big)
\Big\| \leq c\max(\|\tilde{\vh}\|_2^2, \|\tilde{\vm}\|_2^2)\sqrt{(k+n)\log^2 m}.
\end{align}
We have all the ingredients for the final bound on $\mathfrak{C}(\setQ)$ stated below
\begin{align}\label{eq:complexity-estimate}
\mathfrak{C}(\setQ) \leq c\max(\|\tilde{\vh}\|_2^2, \|\tilde{\vm}\|_2^2)\sqrt{(k+n)\log^2 m}. 
\end{align}

\subsection{Probability $p_{\tau}(\setQ)$}
In this section, we determine the probability $p_{\tau}(\setQ)$, and the positive parameter $\tau$ in \eqref{eq:p_tau} for the set $\setQ$ in \eqref{eq:setQ}. For a point $(\dH,\dM) \in \setQ$, and randomly chosen $\nabla \tilde{f}_\ell$, we have via Paley Zygmund inequality that 
\begin{align*}
&\mathbb{P}\Big( \left|\big\<\nabla \tilde{f}_\ell ,(\dH,\dM)\big\> \right|^2 \geq \frac{1}{2} \E\left|\big\<\nabla \tilde{f}_\ell,(\dH,\dM)\big\>  \right|^2 \Big)\geq \frac{1}{4} \frac{\big(\E\big| \big\<\nabla \tilde{f}_\ell ,(\dH,\dM)\big\>  \big|^2\big)^2}{\E\big| \big\<\nabla \tilde{f}_\ell,(\dH,\dM)\big\>  \big|^4}.
\end{align*}
The particular choice of random gradient vectors we are using is $\nabla \tilde{f}_\ell = (1/\sqrt{m}) (|\cl^*\tilde{\vm}|^2 \bl\blt, |\bl^*\tilde{\vh}|^2 \cl\clt)$ giving us $\big\<\nabla \tilde{f}_\ell ,(\dH,\dM)\big\> = (1/\sqrt{m})|\cl^*\tilde{\vm}|^2\<\bl\blt, \dH\> + |\bl^*\tilde{\vh}|^2 \<\cl\clt,\dM\>$. Since $\bl$, and $\cl$ are standard Gaussian vectors, using the equivalence of $L_p$-norms for Gaussians, we deduce that
\begin{align*}
&\left(\E\left| |\cl^*\tilde{\vm}|^2\<\bl\blt, \dH\> +|\bl^*\tilde{\vh}|^2\<\cl\clt,\dM\> \right|^4\right)^{1/4}  \leq \\
&\qquad  c \left(\E\left| |\cl^*\tilde{\vm}|^2\<\bl\blt, \dH\> + |\bl^*\tilde{\vh}|^2 \<\cl\clt,\dM\> \right|^2\right)^{1/2}.
\end{align*}
Plugging last two inequalities in \eqref{eq:p_tau} reveals that
\begin{align}\label{eq:pQ-estimate}
p_{\tau}(\setQ) \geq c > 0
\end{align}
for an absolute constant $c$. To compute $\tau$, we expand $\E\left|\left\<\vg_\ell,(\dH,\dM)\right\>  \right|^2$ giving us
\begin{align}\label{eq:tau-estimate-1}
&\E\left| |\cl^*\tilde{\vm}|^2\<\bl\blt, \dH\> + |\bl^*\tilde{\vh}|^2\<\cl\clt,\dM\> \right|^2 
= 3\|\tilde{\vm}\|_2^4 (\<\text{diag}(\dH),\dH\> + 2\|\dH\|_F^2) \notag\\
&\qquad\qquad +3 \|\tilde{\vh}\|_2^4(\<\text{diag}(\dM),\dM\> + 2\|\dM\|_F^2) + 2|\tilde{\vh}^*\text{diag}(\dH)\tilde{\vh} + 2\tilde{\vh}^*\dH\tilde{\vh}|^2,
\end{align}
where we have made use of multiple simple facts including that $\E |\bl^*\tilde{\vh}|^4 = 3 \|\tilde{\vh}\|_2^4$, and similarly for $|\cl^*\tilde{\vm}|^2$, and two identities: $
\E|\blt\tilde{\vh}|^2 \blt\dH\bl = \tilde{\vh}^* \text{diag}(\dH)\tilde{\vh} + 2\tilde{\vh}^*\dH\tilde{\vh},$
and $\E(\blt\dH\bl) \bl\blt = \text{diag}(\dH) + 2(\dH) \implies \E|\blt\dH\bl|^2 = \<\text{diag}(\dH),\dH\> + 2\|\dH\|_F^2.$
We also made use of the fact that $\setQ \perp \setN$ and therefore $\<\tilde{\mH},\dH\>-\<\tilde{\mM},\dM\> = 0 $, or equivalently, $\tilde{\vh}^*\dH\tilde{\vh} = \tilde{\vm}^*\dM\tilde{\vm}$. 

It is easy to conclude from \eqref{eq:tau-estimate-1} now that 
\begin{align*}
&\E\left| |\cl^*\tilde{\vm}|^2\<\bl\blt, \dH\> + |\bl^*\tilde{\vh}|^2 \<\cl\clt,\dM\> \right|^2  
\geq 6(\|\tilde{\vh}\|_2^4 \|\dH\|_F^2 + \|\tilde{\vm}\|_2^4 \|\dM\|_F^2)\\
&\geq c\min(\|\tilde{\vh}\|_2^2,\|\tilde{\vm}\|_2^2) (\|\dH\|_F^2+\|\dM\|_F^2)
=  c\max(\|\tilde{\vh}\|_2^2,\|\tilde{\vm}\|_2^2) (\|\dH\|_F^2+\|\dM\|_F^2),
\end{align*}
where the last equality uses the fact that $\trace(\tilde{\mH}) = \trace(\tilde{\mM})$ from \eqref{eq:scaled-solution}, which is equivalent to $\|\tilde{\vh}\|_2^2 = \tilde{\vm}\|_2^2$. 
This directly means, we can take $\tau = c \max(\|\tilde{\vh}\|_2^2,\|\tilde{\vm}\|_2^2)$, where $c$ is an absolute constant. 

The complexity estimate in \eqref{eq:complexity-estimate}, value of $\tau$ computed above, and $p_{\tau}(\setQ)$ stated in \eqref{eq:pQ-estimate} together with an application of Lemma \ref{lem:Sample-Complexity} prove Theorem \ref{thm:stable-recovery}.

\section{Proof of Lemma \ref{lem:Sample-Complexity}}
The proof is based on small ball method developed in \cite{koltchinskii2015bounding,mendelson2014learning} and further studied in  \cite{lecue2018regularization} and \cite{lecue2017regularization}. 


Introduce a one sided loss function:
\begin{align}\label{eq:loss}
\setL(\mH,\mM) := \frac{1}{m}\sum_{\ell=1}^m \left[ f_\ell(\mH,\mM)  \right]_+,
\end{align}
where $(\cdot)_+$ denotes the positive side, and $f_\ell(\mH,\mM)$ is a convex function as defined in \eqref{eq:convex-function-for-constraints}. Using this definition, we rewrite \eqref{eq:modified-convex-program} compactly as
\begin{align}\label{eq:cummulative-constr.-loss}
&\minimize~ \setJ(\mH,\mM)\\  
&\text{subject to} ~ \setL(\mH,\mM) \leq 0.\notag
\end{align}
Our objective is to show that any feasible perturbation $(\dH,\dM) \in \setQ$ around any member $(\mH^\nmid,\mM^\nmid)$ of the linearized global scaling set $(\tilde{\mH},\tilde{\mM})\oplus \setN$ has a small Frobenius norm. Feasibility of the perturbation implies that
\begin{align}\label{eq:feasible-descent}
\setL\big(\mH^\nmid+\dH,\mM^\nmid+\dM\big) \leq 0.
\end{align}
Expand the summands $[f_\ell (\mH^\nmid+\dH,\mM^\nmid+\dM)]_+$ of the loss function $\setL\big(\mH^\nmid+\dH,\mM^\nmid+\dM\big)$ to obtain
\begin{align*}
& [f_\ell (\mH^\nmid+\dH,\mM^\nmid+\dM)]_+  =\\
&\gamma_\ell(\mH^\nmid+\dH,\mM^\nmid+\dM) \Bigg[ \sqrt{4\yl^2-\frac{1}{m}\big(\<\bl\blt,\mH^\nmid+\dH\>-\<\cl\clt,\mM^\nmid + \dM\> \big)^2}\\ 
&\qquad\qquad\qquad \qquad -\frac{1}{\sqrt{m}}\big(\<\bl\blt,\mH^\nmid+\dH\>+\<\cl\clt,\mM^\nmid + \dM\> \big)\Bigg]_+.
\end{align*}
Recall that the noisy measurements $y_\ell^2$, defined in \eqref{eq:noisy-measurements}, are related to the noiseless measurements $\tilde{y}_\ell^2$ through $y_\ell^2 = \tilde{y}_\ell^2(1+\xi_\ell)$. Using this relation together with triangle inequality gives
\begin{align}\label{eq:constraint-lowerbound}
&[f_\ell (\mH^\nmid+\dH,\mM^\nmid+\dM)]_+\notag \\
& =\gamma_\ell(\mH^\nmid+\dH,\mM^\nmid+\dM) \Big[ \sqrt{4\tilde{\yl}^2-\tfrac{1}{m}\big(\<\bl\blt,\mH^\nmid+\dH\>-\<\cl\clt,\mM^\nmid + \dM\> \big)^2}\notag\\ 
&\qquad\qquad\qquad\qquad-\tfrac{1}{\sqrt{m}}\big(\<\bl\blt,\mH^\nmid+\dH\>+\<\cl\clt,\mM^\nmid + \dM\> \big)- \sqrt{[-4\tilde{y}_\ell^2\xi_\ell]_+}\Big]_+\notag\\
&\qquad\qquad \geq [\tilde{f}_\ell(\mH^\nmid + \dH, \mM^\nmid+ \dM)]_+ - 2\gamma_\ell(\mH^\nmid+\dH,\mM^\nmid+\dM)  \sqrt{[-\tilde{y}^2_\ell \xi_\ell]_+}\notag\\
&\qquad\qquad \geq [\tilde{f}_\ell(\mH^\nmid + \dH, \mM^\nmid+ \dM)]_+ - 2\sqrt{[-\tilde{y}^2_\ell \xi_\ell]_+}\notag\\
&\qquad\qquad \geq [\tilde{f}_\ell(\tilde{\mH} -\beta \tilde{\mH} + \dH, \tilde{\mM} + \beta \tilde{\mM} + \dM)]_+ - 2 \sqrt{[-\tilde{y}^2_\ell \xi_\ell]_+}\notag\\
&\qquad\qquad\geq [\<\nabla \tilde{f}_\ell, (-\beta\tilde{\mH} +\dH, \beta \tilde{\mM}+\dM)\>]_+-2\sqrt{[-\tilde{y}^2_\ell \xi_\ell]_+} \notag\\
&\qquad\qquad= [\<\nabla \tilde{f}_\ell, (-\beta\tilde{\mH} +\dH, \beta \tilde{\mM}+\dM)\>]_+ - 2\sqrt{[-\tilde{y}^2_\ell \xi_\ell]_+}\notag\\
&\qquad\qquad = [\<\nabla \tilde{f}_\ell, (\dH,\dM)\>]_+ - 2\sqrt{[-\tilde{y}^2_\ell \xi_\ell]_+},
\end{align}
where in the first inequality follows from the fact that if $a \geq 0$, and $b < 0$ with $a+b \geq 0$, then $\sqrt{a+b} \geq \sqrt{a}- \sqrt{-b}$ holds, and if $a \geq 0$, and $b \geq 0$, $\sqrt{a+b} \geq \sqrt{a}$ holds,  the second inequality uses the fact that $\tilde{f}_\ell(\mH^\nmid+\dH,\mM^\nmid+\dM) \leq 0$ as $(\dH,\dM)$ is a feasible perturbation of $(\mH^\nmid,\mM^\nmid)$, and hence using the definition \eqref{eq:gamma-def} it holds that $\gamma_\ell(\mH^\nmid+\dH,\mM^\nmid+\dM) \leq 1$, the third simply uses the fact that $(\mH^\nmid, \mM^\nmid) \in (\tilde{\mH},\tilde{\mM}) \oplus \setN$ is of the form $(\mH^\nmid, \mM^\nmid) = ((1-\beta)\tilde{\mH},(1+\beta)\tilde{\mM})$ for some $\beta \in \R$ (More precisely, the scalar $\beta \in [-1,1]$, as by feasibility $\mH^\nmid$, and $\mM^\nmid$ are PSD), and finally the last inequality uses the definition of sub-gradient \eqref{eq:gradient-ftilde} of the convex function $\tilde{f}_\ell$. The last equality uses the fact that $\<\nabla \tilde{f}_\ell, (-\tilde{\mH},\tilde{\mM})\> = 0$. 

Plugging the lower bound \eqref{eq:constraint-lowerbound} in \eqref{eq:feasible-descent} produces
\begin{align}\label{eq:lower-bound-loss}
\sum_{\ell=1}^m [\<\nabla \tilde{f}_\ell, (\dH,\dM)\> ]_+ &\leq 2\sum_{\ell=1}^m \sqrt{|\tilde{y}^2_\ell \xi_\ell|} \leq \frac{2}{\sqrt{m}} \sqrt{\|\vxi\|_\infty }\sum_{\ell=1}^m |\bl^*\tilde{\vh}||\cl^*\tilde{\vm}|\notag\\
 &\leq 2\sqrt{\frac{\|\vxi\|_\infty}{m}} \|\mB\tilde{\vh}\|_2 \|\mC \tilde{\vm}\|_2 \leq 18  \sqrt{m\|\vxi\|_\infty} \|\tilde{\vh}\|_2\|\tilde{\vm}\|_2,
\end{align}
where the second last display follows just by using the fact that $\tilde{y}_\ell^2 = \frac{1}{m} \<\bl\bl^*,\tilde{\mH}\>\<\cl\cl^*,\tilde{\mM}\> $, where $\tilde{\mH} = \tilde{\vh}\tilde{\vh}^*$, and $\tilde{\mM} = \tilde{\vm}\tilde{\vm}^*$, and the last display simply employs Cauchy Schwarz, and $\|\mB\|\leq 3\sqrt{m}$, and $\|\mC\|\leq 3 \sqrt{m}$, which holds with probability at least $1-\mathrm{e}^{-m/2}$. 

Let $\psi_t(s) := (s)_+-(s-t)_+$. Using the fact that  $\psi_t(s) \leq (s)_+$, and that for every $\alpha, t \geq 0$, and $s \in \R$, $\psi_{\alpha t}(s) = t\psi_{\alpha}(\frac{s}{t})$, we have
\begin{align}\label{eq:interim-eq2}
\frac{1}{\sqrt{m}}\sum_{\ell=1}^m\big[\big\<\nabla \tilde{f}_\ell,(\dH,\dM)\big\>\big]_+&\geq \frac{1}{\sqrt{m}}\sum_{\ell=1}^m\psi_{\tau \|(\dH,\dM)\|_F}(\big\<\nabla \tilde{f}_\ell,(\dH,\dM)\big\>)\notag\\
& = \|(\dH,\dM)\|_F\cdot \frac{1}{\sqrt{m}}\sum_{\ell=1}^m\psi_{\tau}\big(\big\<\nabla \tilde{f}_\ell ,\tfrac{(\dH,\dM)}{\|(\dH,\dM)\|_F}\big\>\big)\notag\\
&= \|(\dH,\dM)\|_F\cdot \frac{1}{\sqrt{m}}\Bigg[\sum_{\ell=1}^m\E \psi_{\tau}\big(\big\<\nabla \tilde{f}_\ell,\tfrac{(\dH,\dM)}{\|(\dH,\dM)\|_F}\big\>\big) - \notag\\
& \sum_{\ell=1}^m \bigg[ \E \psi_{\tau}\big(\big\<\nabla \tilde{f}_\ell,\tfrac{(\dH,\dM)}{\|(\dH,\dM)\|_F}\big\>\big)- \psi_{\tau}\big(\big\<\nabla \tilde{f}_\ell,\tfrac{(\dH,\dM)}{\|(\dH,\dM)\|_F}\big\>\big) \bigg]\Bigg].
\end{align}
Define a centered random process $\mathcal{R}(\mB,\mC)$ as follows
\begin{align*}
&\mathcal{R}(\mB,\mC):=\sup_{(\dH,\dM)\in \setQ}\frac{1}{\sqrt{m}}\sum_{\ell=1}^m\bigg[\E \psi_{\tau}\big(\big\<\nabla \tilde{f}_\ell,\tfrac{(\dH,\dM)}{\|(\dH,\dM)\|_F}\big\>\big)- \psi_{\tau}\big(\big\<\nabla \tilde{f}_\ell,\tfrac{(\dH,\dM)}{\|(\dH,\dM)\|_F}\big\> \big)\bigg]
\end{align*}
and an application of bounded difference inequality \cite{mcdiarmid1989method} yields that $\mathcal{R}(\mB,\mC) \leq \E \mathcal{R}(\mB,\mC) + t\tau/\sqrt{m}$  with probability at least $1-\mathrm{e}^{-2mt^2}$. It remains to evaluate $\E  \mathcal{R}(\mB,\mC)$, which after using a simple symmetrization inequality \cite{van1997weak} yields 
\begin{align}
&\E \setR(\mB,\mC) \leq 2\E \sup_{(\dH,\dM)\in \setQ}\frac{1}{\sqrt{m}}\sum_{\ell=1}^m\varepsilon_\ell \psi_{\tau}\big(\big\<\nabla \tilde{f}_\ell,\tfrac{(\dH,\dM)}{\|(\dH,\dM)\|_F}\big\>\big),
\end{align}
where $\varepsilon_1, \varepsilon_2, \ldots, \varepsilon_m$ are independent Rademacher random variables. Using the fact that $\psi_t(0) = 0$, and $\psi_t(s)$ is a contraction: $|\psi_t(\alpha_1)-\psi_t(\alpha_2)| \leq |\alpha_1-\alpha_2|$ for all $\alpha_1, \alpha_2 \in \R$, we have from the Rademacher contraction inequality \cite{ledoux2013probability} that 
\begin{align}\label{eq:random-process}
&\E \sup_{(\dH,\dM)\in \setQ}\frac{1}{\sqrt{m}}\sum_{\ell=1}^m\varepsilon_\ell \psi_{\tau}\big(\big\<\nabla \tilde{f}_\ell,\tfrac{(\dH,\dM)}{\|(\dH,\dM)\|_F}\big\>\big)\leq \E \sup_{(\dH,\dM)\in \setQ}\frac{1}{\sqrt{m}}\sum_{\ell=1}^m\varepsilon_\ell\big\<\nabla \tilde{f}_\ell,\tfrac{(\dH,\dM)}{\|(\dH,\dM)\|_F}\big\>\notag\\
&\qquad\qquad = \E \sup_{(\dH,\dM)\in \setQ}\frac{1}{\sqrt{m}}\sum_{\ell=1}^m\varepsilon_\ell\big\<\nabla \tilde{f}_\ell,\tfrac{(\dH,\dM)}{\|(\dH,\dM)\|_F}\big\>, 
\end{align}
where the last equality is the result of the fact that a global sign change of a sequence of  Rademacher random variables does not change their distribution. In addition, using the facts that $t\mathbf{1}(s\geq t) \leq \psi_t(s)$, and that random vectors $\nabla \tilde{f}_1, \nabla \tilde{f}_2, \ldots, \nabla \tilde{f}_m$ are identically distributed and the distribution is symmetric, it follows 
\begin{align}\label{eq:tail-prob}
&\frac{\tau}{\sqrt{m}}\PP\big(\big\<\nabla \tilde{f}_\ell,\tfrac{(\dH,\dM)}{\|(\dH,\dM)\|_F}\big\>\geq \frac{\tau}{\sqrt{m}}\big) = \frac{\tau}{\sqrt{m}}\E \bigg(\mathbf{1}\bigg[{ \big\<\nabla \tilde{f}_\ell,\tfrac{(\dH,\dM)}{\|(\dH,\dM)\|_F}\big\>\geq \frac{\tau}{\sqrt{m}}}\bigg]\bigg)\notag\\
&\leq \E \psi_{\tau/\sqrt{m}}\left(\big\<\nabla \tilde{f}_\ell,\tfrac{(\dH,\dM)}{\|(\dH,\dM)\|_F}\big\>\right) = \frac{1}{\sqrt{m}}\E \psi_{\tau}\left(\big\<\nabla \tilde{f}_\ell,\tfrac{(\dH,\dM)}{\|(\dH,\dM)\|_F}\big\>\right). 
\end{align}
Plugging \eqref{eq:tail-prob}, and \eqref{eq:random-process} in \eqref{eq:interim-eq2}, we have 
\begin{align*}
&\frac{1}{\sqrt{m}}\sum_{\ell=1}^m\big[\big\<\nabla \tilde{f}_\ell,(\dH,\dM)\big\>\big]_+\geq \tau\|(\dH,\dM)\|_F\cdot\PP\big(\big\<\nabla \tilde{f}_\ell,\tfrac{(\dH,\dM)}{\|(\dH,\dM)\|_F}\big\>\geq \frac{\tau}{\sqrt{m}}\big)  \\
&-2\|(\dH,\dM)\|_F \E \sup_{(\dH,\dM)\in \setQ} \frac{1}{\sqrt{m}}\sum_{\ell=1}^m\varepsilon_\ell\big\<\nabla \tilde{f}_\ell,\tfrac{(\dH,\dM)}{\|(\dH,\dM)\|_F}\big\>-2\|(\dH,\dM)\|_F\frac{t\tau}{\sqrt{m}}.\\
\end{align*}
Using this lower bound in \eqref{eq:lower-bound-loss}, we obtain 
\begin{align*}
&\|(\dH,\dM)\|_F\Big[\tau \PP\big(\big\<\nabla \tilde{f}_\ell,\tfrac{(\dH,\dM)}{\|(\dH,\dM)\|_F}\big\>\geq \frac{\tau}{\sqrt{m}}\big)-2\E \sup_{(\dH,\dM)\in \setQ} \frac{1}{\sqrt{m}}\sum_{\ell=1}^m\varepsilon_\ell\big\<\nabla\tilde{f}_\ell,\tfrac{(\dH,\dM)}{\|(\dH,\dM)\|_F}\big\>\Big]\\
&\qquad\qquad -2\|(\dH,\dM)\|_F\frac{t\tau}{\sqrt{m}} \leq 18  \sqrt{\|\vxi\|_\infty} \|\tilde{\vh}\|_2\|\tilde{\vm}\|_2.
\end{align*}
Using the definitions in \eqref{eq:Rademacher-Complexity}, and \eqref{eq:p_tau}, we can write 
\begin{align*}
\|(\dH,\dM)\|_F \left(\tau p_{\tau}(\setQ)- \frac{(2\mathfrak{C}(\setQ) + t\tau)}{\sqrt{m}}\right) \leq 18  \sqrt{\|\vxi\|_\infty} \|\tilde{\vh}\|_2\|\tilde{\vm}\|_2.
\end{align*}
It is clear that choosing $m \geq \left( \frac{2\mathfrak{C}(\setQ)+t\tau}{\tau p_\tau(\setQ)}\right)^2$ implies that any feasible direction $(\dH,\dM) \in \setN_\perp$ is bounded by 
\begin{align}\label{eq:feasible-direction-bound}
\|(\dH,\dM)\|^2_F \leq 18^2  \|\vxi\|_\infty \|\tilde{\mH}\|_F\|\tilde{\mM}\|_F
\end{align}
with probability at least $1-\mathrm{e}^{-c_t m}$, where $c_t = ct^2$ for a universal constant $c$, where we used the fact that $\tilde{\mH} = \tilde{\vh}\tilde{\vh}^*$, and $\tilde{\mM} = \tilde{\vm}\tilde{\vm}^*$. Since $(\dH,\dM) \in \setN_\perp$, the last display only gives us that an element, $((1-\beta_0)\tilde{\mH},(1+\beta_0)\tilde{\mM})$ for some $\beta_0 \in \R$, of the set $(\tilde{\mH},\tilde{\mM}) \oplus \setN$ obeys
\begin{align}\label{eq:feasible-direction-bound-1}
\|(\widehat{\mH},\widehat{\mM}) - ((1-\beta_0)\tilde{\mH},(1+\beta_0)\tilde{\mM}) \|_F^2\leq 18^2  \|\vxi\|_\infty \|\tilde{\mH}\|_F\|\tilde{\mM}\|_F. 
\end{align}
That is, the solution $(\widehat{\mH},\widehat{\mM})$ cannot wander too far away from the line $(\tilde{\mH},\tilde{\mM}) \oplus \setN$. We call this norm cylinder constraint as the solution must lie within a cylinder, centered at a line $(\tilde{\mH},\tilde{\mM}) \oplus \setN$ and of radius given by the rhs of the last display above.  Equivalently, a displacement $((1-\beta_0)\tilde{\mH},(1+\beta_0)\tilde{\mM})$  of the ground truth $(\tilde{\mH}, \tilde{\mM})$ is sufficiently close to $(\widehat{\mH},\widehat{\mM})$. Using this fact together with the fact that the feasible hyperbolic set diverges away from the $(\tilde{\mH},\tilde{\mM}) \oplus \setN$ for larger displacement $\beta$, we will conclude in the remaining proof that the displacement $\beta_0$ cannot be too large, and hence the Euclidean distance between $(\widehat{\mH},\widehat{\mM})$, and the ground truth $(\tilde{\mH},\tilde{\mM})$ is also bounded. 
\\
\\
\noindent \textbf{Case 1:} Assume that noise $\vxi$ is such that $\xi_\ell \in [-1,0]$ for every $\ell \in [m]$, and $\exists ~ \ell^\prime \in [m]$ such that $\xi_{\ell^\prime}= 0$. 

Trivially, the minimizer $(\widehat{\mH},\widehat{\mM})$ must lie somewhere in the feasible set specified by the $\ell^\prime$ constraint: $\tfrac{1}{m}\<\vb_{\ell^\prime}\vb_{\ell^\prime}^*,\mH\>\<\vc_{\ell^\prime}\vc_{\ell^\prime}^*,\mM\> \geq  \tilde{y}_{\ell^\prime}^2$. Define the boundary $\setB$ of the feasible set above as follows
\begin{align}\label{eq:hyperbolic-set}
\setB := \{(\mH,\mM): \tfrac{1}{m}\<\vb_{\ell^\prime}\vb_{\ell^\prime}^*,\mH\>\<\vc_{\ell^\prime}\vc_{\ell^\prime}^*,\mM\> =  \tilde{y}_{\ell^\prime}^2\}.
\end{align}
The line $(\tilde{\mH}, \tilde{\mM}) \oplus \setN$ only touches the feasible set at $(\tilde{\mH}, \tilde{\mM})$. Define a plane $\setP := \text{span}\{(\tilde{\mH},\mathbf{0}), (\mathbf{0},\tilde{\mM})\}$.  Clearly, the line $(\tilde{\mH},\tilde{\mM}) \oplus \setN$ is contained in $\setP$. Moreover, the intersection $\setP \cap \setB$ only happens at a set of bilinear scaling ambiguity, that is, $\setP\cap\setB = \{(\gamma \tilde{\mH}, \frac{1}{\gamma} \tilde{\mM})~ \text{for some} ~~\gamma \in \R -\{0\}\}$.  Observe that out of all fixed norm feasible points, the one that leads to a largest displacement $\beta_0$ must be on the hyperbolic set $\setP\cap\setB$. Use this fact to conclude that in the worst case (largest $\beta_0$) solution point $(\widehat{\mH}, \widehat{\mM})$ must be equal to $(\gamma \tilde{\mH}, \tfrac{1}{\gamma}\tilde{\mM})$ for some $\gamma \in \R-\{0\}$.  In general, the Euclidean distance between a point $(\gamma \tilde{\mH},\tfrac{1}{\gamma}\tilde{\mM}) \in \setP \cap \setB $ and its orthogonal projection  $((1-\beta)\tilde{\mH},(1+\beta)\tilde{\mM})$ onto the line $(\tilde{\mH},\tilde{\mM})\oplus \setN$ is given by
\begin{align}\label{eq:distance}
&\|(\gamma \tilde{\mH},\tfrac{1}{\gamma}\tilde{\mM}) - ((1-\beta)\tilde{\mH},(1+\beta)\tilde{\mM}) \|_F^2 =\notag\\
&\qquad\qquad  \left[\left(\frac{2\beta + \sqrt{\beta^2+2}}{\sqrt{2}}-1\right)^2+ \left(\frac{-2\beta + \sqrt{\beta^2+2}}{\sqrt{2}}-1\right)^2\right]\|(\tilde{\mH}, \tilde{\mM})\|_F^2.
\end{align}
In light of \eqref{eq:feasible-direction-bound-1}, we then have 
\begin{align*}
&\left[\left(\frac{2\beta_0 + \sqrt{\beta_0^2+2}}{\sqrt{2}}-1\right)^2+ \left(\frac{-2\beta_0 + \sqrt{\beta_0^2+2}}{\sqrt{2}}-1\right)^2\right]\|(\tilde{\mH}, \tilde{\mM})\|_F^2 \leq 18^2  \|\vxi\|_\infty \|\tilde{\mH}\|_F\|\tilde{\mM}\|_F\\
& \qquad\qquad \implies 4\beta_0^2\|(\tilde{\mH}, \tilde{\mM})\|_F^2 \leq 18^2  \|\vxi\|_\infty \|\tilde{\mH}\|_F\|\tilde{\mM}\|_F, 
\end{align*}
where the implication follows by using the fact that 
\begin{align*}
\left(\frac{2\beta + \sqrt{\beta^2+2}}{\sqrt{2}}-1\right)^2+ \left(\frac{-2\beta + \sqrt{\beta^2+2}}{\sqrt{2}}-1\right)^2 \geq 4 \beta^2
\end{align*}
holds for any $\beta \in \R$. Using the bound on $\beta_0$ developed above, the conclusion in \eqref{eq:feasible-direction-bound-1} can be refined using trangle inequality to 
\begin{align*}
\|(\widehat{\mH},\widehat{\mM}) - (\tilde{\mH},\tilde{\mM}) \|_F - \beta_0\|(\tilde{\mH},\tilde{\mM})\|_F &\leq 18  \sqrt{\|\vxi\|_\infty \|\tilde{\mH}\|_F\|\tilde{\mM}\|_F}\\
\|(\widehat{\mH},\widehat{\mM}) - (\tilde{\mH},\tilde{\mM}) \|_F^2 &\leq 405 \|\vxi\|_\infty \|\tilde{\mH}\|_F\|\tilde{\mM}\|_F.
\end{align*}

\noindent\textbf{Case 2:} We now consider general case of noise when $\vxi$ is such that $\xi_\ell \geq -1$.

The key idea is that the measurements with noise as in Case 2 can be converted to the measurements with noise as in Case 1. To see this, define
\begin{align*}
s := \max_{\ell \in [m]} \frac{y_\ell}{\tilde{y}_\ell} = 1+ \|\vxi\|_\infty, \ 
 \eta_\ell := \frac{1}{s}(1-s+\xi_\ell). 
\end{align*}
From the definitions above, it can be easily verified that $s = 1+ \|\vxi\|_\infty$, and that $(1+\xi_\ell) = s(1+\eta_\ell)$. Using this relation allows us to rewrite measurements $y_\ell = \tilde{y}_\ell(1+\xi_\ell)$ contaminated with noise $\xi_\ell$ equivalently as $y_\ell  = s\tilde{y}_\ell (1+\eta_\ell)$, where $\eta_\ell$ is now interpreted as noise, and the new scaled noiseless measurements are interpreted as $sy_\ell$. We will now show that $\eta_\ell \in [-1,0]$.

 By definition, $s\geq 1+\xi_\ell$, this implies that $\eta_\ell \leq (1/s)(s-s) = 0$. Also note that $\eta_\ell = 0$ for a $\xi_\ell$ that achieves maximum $\|\vxi\|_\infty$.  Moreover, using the definition of $\eta_\ell$, and the fact that $\xi_\ell \geq -1$ gives $\eta_\ell \geq -1$. 
 
 Firstly, the new noise $\eta_\ell$ obeys all the conditions in Case 1 above. Secondly, since the noiseless measurements $\tilde{y}_\ell^2$ of $(\tilde{\mH},\tilde{\mM})$ are $\tilde{y}_\ell^2 = \tfrac{1}{m}\<\bl\bl^*,\tilde{\mH}\>\<\cl\cl^*,\tilde{\mM}\>$, we can interpret $s\tilde{y}_\ell^2$ as the noiseless measurements of $(\sqrt{s}\tilde{\mH},\sqrt{s}\tilde{\mM})$. We can now directly invoke result of Case 1 here to obtain 
 \begin{align}\label{eq:interim}
 &\|(\widehat{\mH},\widehat{\mM}) - (\sqrt{s}\tilde{\mH},\sqrt{s}\tilde{\mM})\|^2_F = \|\widehat{\mH} - \sqrt{s}\tilde{\mH} \|_F^2 + \|\widehat{\mM} - \sqrt{s}\tilde{\mM} \|_F^2 \notag \\
 & \leq 405  \|\boldsymbol{\eta}\|_\infty^2 \|\sqrt{s}\tilde{\mH}\|_F\|\sqrt{s}\tilde{\mM}\|_F \leq 405 s  \|\boldsymbol{\eta}\|_\infty \|\tilde{\mH}\|_F\|\tilde{\mM}\|_F \leq 810 \|\vxi\|_\infty \|\tilde{\mH}\|_F\|\tilde{\mM}\|_F,
 \end{align}
 where the last inequality is obtained  using the inequality derived below 
 \begin{align*}
 & s\eta_\ell = 1-s + \xi_\ell = \xi_\ell - \|\vxi\|_\infty \implies s |\eta_\ell| = \|\vxi\|_\infty - \xi_\ell \\
 & \implies  s\|\boldsymbol{\eta}\|_\infty = \|\vxi\|_\infty - \min_{\ell \in [m]} \xi_\ell \leq 2 \|\vxi\|_\infty.
 \end{align*}
Observe that 
\begin{align*}
&\|(\widehat{\mH},\widehat{\mM}) - (\tilde{\mH},\tilde{\mM})\|_F = \sqrt{\|\widehat{\mH} - \tilde{\mH} \|_F^2 + \|\widehat{\mM} - \tilde{\mM} \|_F^2 }\\
& = \sqrt{\|\widehat{\mH} - \sqrt{s}\tilde{\mH}+\sqrt{s}\tilde{\mH} - \tilde{\mH} \|_F^2 + \|\widehat{\mM} -\sqrt{s}\tilde{\mM}+\sqrt{s}\tilde{\mM} - \tilde{\mM} \|_F^2}\\
& \leq 2\left(\sqrt{\|\widehat{\mH} - \sqrt{s}\tilde{\mH}\|_F^2 +  \|\widehat{\mM} - \sqrt{s}\tilde{\mM}\|_F^2 }+ (\sqrt{s}-1) \sqrt{\|\tilde{\mH}\|_F^2 + \|\tilde{\mM}\|_F^2}\right)\\
& \leq 2\sqrt{810} \sqrt{\|\vxi\|_\infty} \sqrt{\|\tilde{\mH}\|_F\|\tilde{\mM}\|_F} + 2(\sqrt{s}-1)\sqrt{\|\tilde{\mH}\|_F^2+\|\tilde{\mM}\|_F^2} \\
& \leq 44 \sqrt{\|\vxi\|_\infty}\sqrt{\|\tilde{\mH}\|_F^2+\|\tilde{\mM}\|_F^2},
\end{align*}
where the second last inequality follows by using \eqref{eq:interim}. 
The proof is complete.



\section*{Appendix A. Proof of Lemma \ref{lem:convexity}}\label{appx:convexity}
	The objective of \eqref{eq:convex-optimization-program} is simply linear, we focus on the constraints. For a fixed $\ell$, let $S_{\ell} := \{(\mH,\mM) \in \setH^{k \times k} \times \setH^{m \times m} \ | \ \tfrac{1}{m} \<\bl\bl^*,\mH\>\<\cl\cl^*,\mM\> \geq \tilde{y}^2_\ell, \mH \succcurlyeq \vzero, \mM \succcurlyeq \vzero\},$ $S_{\ell,1} := \{(u_\ell, v_\ell) \in \R^2\ | \ \tfrac{1}{m} u_\ell v_\ell \geq \tilde{y}^2_\ell, u_\ell \geq 0, v_\ell \geq 0\},$ and $S_{\ell,2} := \{(\mH,\mM) \in \setH^{k \times k} \times \setH^{m \times m} \ | \ (\<\bl\blt, \mH\>,\<\cl\clt,\mM\>) \in S_{\ell,1}\}$. To show that $S_\ell$ is convex, it suffices to show that $S_{\ell,1}$, and $S_{\ell,2}$ are convex. 
	
	Fix $(u_1,v_1), (u_2,v_2) \in S_{\ell,1}$, and let $\alpha \in [0,1]$. Note that $u_1 > 0$, and $u_2 > 0$ as $y_\ell > 0$. Consider 
	\begin{align*}
	&\frac{1}{m}(\alpha u_1 + (1-\alpha)u_2)  (\alpha v_1 + (1-\alpha)v_2)\\
	&\qquad\qquad=  \frac{1}{m}\left((\alpha^2 u_1v_1 + (1-\alpha)^2u_2v_2) + \alpha(1-\alpha) (u_1v_2+u_2v_1)\right)\\
	& \qquad\qquad \geq (\alpha^2 \tilde{y}_\ell^2 + (1-\alpha)^2\tilde{y}_\ell^2) + \alpha(1-\alpha) (\frac{\tilde{y}_\ell^2 u_1}{u_2} +\frac{\tilde{y}_\ell^2 u_2}{u_1})\\
	&\qquad\qquad = \tilde{y}^2_\ell \left(1 + \frac{2\alpha^2 u_1u_2-2\alpha u_1u_2 + \alpha(1-\alpha) (u_1^2+u_2^2)}{u_1u_2}\right)\\
	&\qquad\qquad= \tilde{y}_\ell^2 \left( 1 + \frac{(\alpha-\alpha^2)(u_1-u_2)^2}{u_1u_2}\right) \geq \tilde{y}_\ell^2,\\
	\end{align*}
	where the last inequality follows form the fact that $\alpha \in [0,1]$, and $u_1u_2 > 0$. This shows that $S_{\ell,1}$ is convex. 
	
	The set $S_{\ell,2}$ is convex as the inverse image of a convex set of a linear map is convex. This implies that $S_\ell$ is convex. Finally, the intersection of any number of convex sets is convex means that the constraint of \eqref{eq:convex-optimization-program} is convex. This proves that \eqref{eq:convex-optimization-program} is a convex program.

\section*{Acknowledgement}
PH acknowledges support from NSF DMS 1464525. 


%

\end{document}